\pdfoutput=1

\documentclass[11pt]{article}

\usepackage[]{acl}

\usepackage{times}
\usepackage{latexsym}

\usepackage[T1]{fontenc}

\usepackage[utf8]{inputenc}

\usepackage{comment}

\usepackage{microtype}

\usepackage{amsmath,amsfonts,bm}

\def\eqref#1{equation~\ref{#1}}

\def\1{\bm{1}}

\def\rmA{{\mathbf{A}}}

\def\rmT{{\mathbf{T}}}

\def\rmV{{\mathbf{V}}}

\def\va{{\bm{a}}}
\def\vb{{\bm{b}}}

\def\vt{{\bm{t}}}

\def\vv{{\bm{v}}}

\DeclareMathAlphabet{\mathsfit}{\encodingdefault}{\sfdefault}{m}{sl}
\SetMathAlphabet{\mathsfit}{bold}{\encodingdefault}{\sfdefault}{bx}{n}

\def\gL{{\mathcal{L}}}

\def\sR{{\mathbb{R}}}

\usepackage{pifont}
\usepackage{tabu}
\usepackage{hyperref}
\usepackage{url}
\usepackage{svg}

\usepackage{xcolor,colortbl}
\usepackage{natbib}
\usepackage{soul}
\usepackage{tikz}
\usepackage{tikz-dependency}
\usetikzlibrary{positioning}
\usetikzlibrary{calc,intersections}
\usepackage[utf8]{inputenc}

\usepackage[nocenter]{qtree}

\usepackage{CJKutf8}

\usepackage{amsmath,amsthm,amssymb,amsfonts,bbm,stmaryrd}
\usepackage{enumitem}

\usepackage{textcomp}
\usepackage{multirow} 
\usepackage{graphicx} 
\usepackage{subcaption}
\usepackage{booktabs} 
\usepackage[normalem]{ulem}
\usepackage{makecell}
\usepackage{courier}
\usepackage{bbm}
\usepackage{algorithm,algpseudocode}
\usepackage{adjustbox}
\usepackage{tabularx}
\usepackage{booktabs}
\usepackage{setspace}

\usepackage{threeparttable}
\usepackage{makecell}
\usepackage{cancel}

\usepackage{array}
\newcolumntype{P}[1]{>{\centering\arraybackslash}p{#1}}
\newcolumntype{M}[1]{>{\centering\arraybackslash}m{#1}}
\newcommand\bothmodels[3]{\multicolumn{#1}{c}{\cellcolor{#3} #2}}
\newcommand\coloredrow[3]{\multicolumn{#1}{l}{\cellcolor{#2} #3}}

\newcommand{\yanpeng}[1]{\textcolor{red}{YZ: #1}}
\newcommand{\youngjae}[1]{\textcolor{orange}{YY: #1}}

\definecolor{myorange}{HTML}{FF6700}
\definecolor{mypurple}{HTML}{9900ff}
\definecolor{mygreen}{HTML}{009000}
\tikzset{state/.style={
		circle,
		minimum size =1.25cm
		draw=black,
		thick,
		fill=orange,
		text=white}}
\tikzset{block/.style={
		rounded corners, fill=blue!15,
		anchor=north west}}
\tikzset{line/.style={
		-latex,
		line width=.5mm,
		black!65}}
\tikzset{annotation/.style={
		align=left,
		anchor=north west,
		draw=black!75,
		minimum height=10mm,
		minimum width=10mm}}
\tikzset{hmodule/.style={
		rounded corners,
		align=center,
		anchor=north west,
		font=\Large,
		fill=blue!20,
		draw=lightgray!75,
		minimum height=10mm,
		minimum width=20mm}}
\tikzset{vmodule/.style={
		rounded corners,
		align=center,
		rotate=90,
		anchor=north west,
		font=\Large,
		fill=blue!20,
		draw=lightgray!75,
		minimum height=10mm,
		minimum width=20mm}}
\tikzset{hvector/.style={
		rounded corners=2.5mm,
		align=center,
		anchor=north west,
		fill=gray!5,
		draw=black!75,
		minimum height=5mm,
		minimum width=20mm}}
\tikzset{vvector/.style={
		rounded corners=2.5mm,
		align=center,
		rotate=90,
		anchor=north west,
		fill=gray!5,
		draw=black!75,
		minimum height=5mm,
		minimum width=20mm}}

\usepackage{arydshln}
\makeatletter
\def\adl@drawiv#1#2#3{%
        \hskip.5\tabcolsep
        \xleaders#3{#2.5\@tempdimb #1{1}#2.5\@tempdimb}%
                #2\z@ plus1fil minus1fil\relax
        \hskip.5\tabcolsep}
\newcommand{\cdashlinelr}[1]{%
  \noalign{\vskip\aboverulesep
           \global\let\@dashdrawstore\adl@draw
           \global\let\adl@draw\adl@drawiv}
  \cdashline{#1}
  \noalign{\global\let\adl@draw\@dashdrawstore
           \vskip\belowrulesep}}
\makeatother

\usepackage{xspace}
\usepackage[T1]{fontenc}    

\newcommand\merlottitlefont[1]{\smash{{\usefont{T1}{QTWestEndRegular}{m}{n}#1}}}
\newcommand\merlotfont[1]{\smash{{\usefont{T1}{QTWestEndRegular}{m}{n}#1}}}

\newcommand{\modelname}{\merlotfont{vip-AnT}\xspace}
\newcommand{\modelnamelong}{
\merlottitlefont{VI}sually \merlottitlefont{P}ivoted \merlottitlefont{A}udio and(\merlottitlefont{N}) \merlottitlefont{T}ext\xspace}

\title{Connecting the Dots between Audio and Text without Parallel Data\\ 
through Visual Knowledge Transfer
}

\author{  
Yanpeng Zhao$^{\clubsuit}$\thanks{\,\,\,Work was partially done during an internship at AI2.} 
\: \:
Jack Hessel$^{\heartsuit}$ \: \:
Youngjae Yu$^{\heartsuit}$ \: \: \\ \bf
Ximing Lu$^{\spadesuit\heartsuit}$ \: \: 
Rowan Zellers$^\spadesuit$ \: \: 
Yejin Choi$^{\spadesuit\heartsuit}$\\
$^\clubsuit$Institute for Language, Cognition and Computation, University of Edinburgh \\
$^\spadesuit$Paul G. Allen School of Computer Science \& Engineering, University of Washington \\
$^\heartsuit$Allen Institute for Artificial Intelligence\\
}

\begin{document}
\maketitle
\begin{abstract}
Machines that can represent and describe environmental soundscapes have practical potential, e.g., for audio tagging and captioning.
Prevailing learning paradigms of audio-text connections have been relying on parallel audio-text data, which is, however, scarcely available on the web.
We propose \modelname that induces \textbf{A}udio-\textbf{T}ext alignment without using any parallel audio-text data.
Our key idea is to share the image modality between bi-modal image-text representations and bi-modal image-audio representations;
the image modality functions as a pivot and connects audio and text in a tri-modal embedding space implicitly.

In a difficult zero-shot setting with no paired audio-text data, our model demonstrates state-of-the-art zero-shot performance on the ESC50 and
US8K audio classification tasks, and even
surpasses the supervised state of the art for Clotho caption retrieval (with audio queries) by 2.2\% R@1.
We further investigate cases of minimal audio-text supervision, finding that, e.g., just a few hundred supervised audio-text pairs increase the zero-shot audio classification accuracy by 8\% on US8K.
However, to match human parity on some zero-shot tasks, our empirical scaling experiments suggest that we would need about $2^{21} \approx 2\text{M}$ supervised audio-caption pairs.
Our work opens up new avenues for learning audio-text connections with little to no parallel audio-text data.

\end{abstract}

\section{Introduction}

\begin{figure}[t!]
  \centering\small
    \includegraphics[width=1.0\columnwidth]{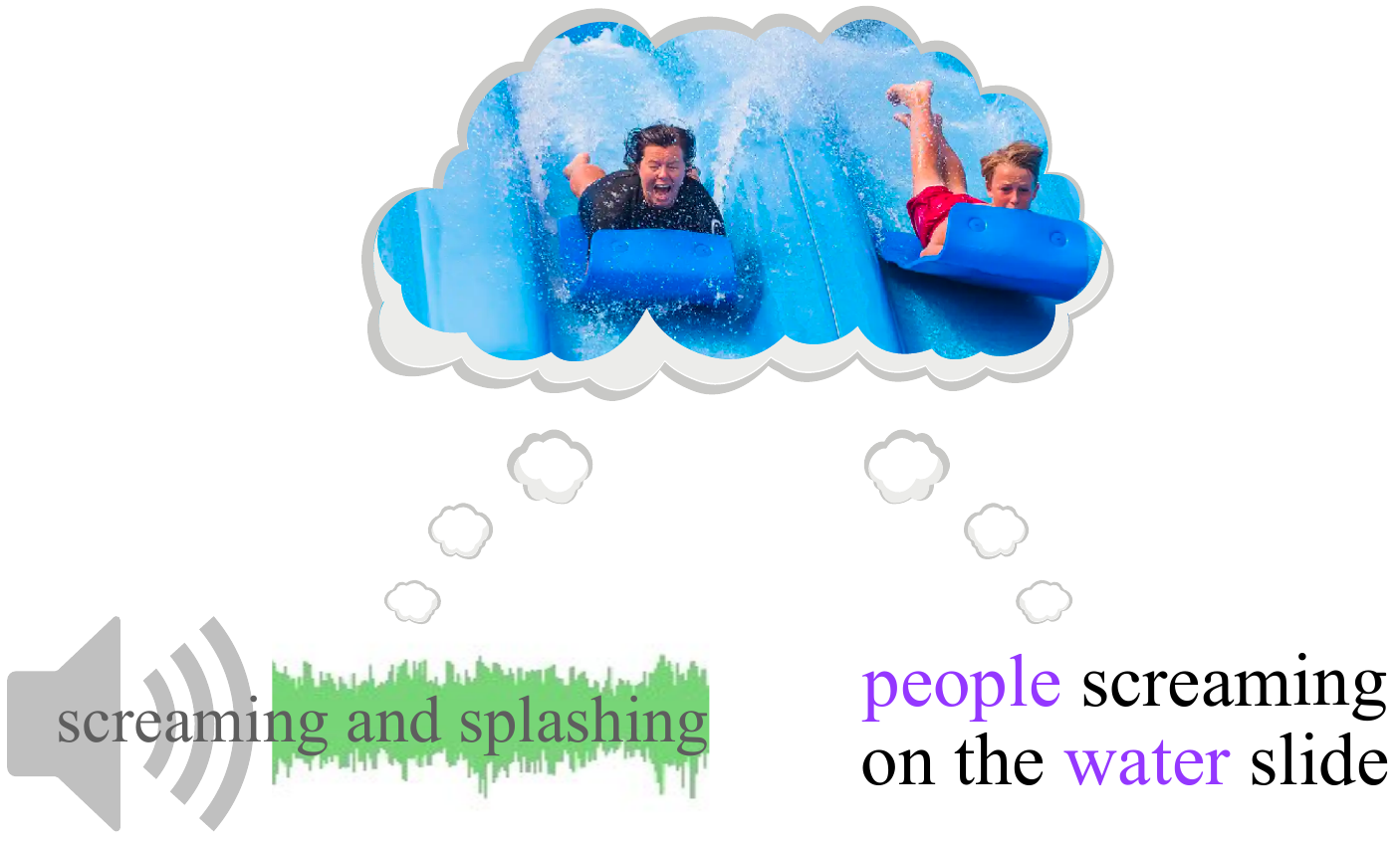}
\caption{\label{fig:image_pivot_idea}
\modelname pivots audio and text via visual imagination.
}
\end{figure}

Environmental sound provides rich perspectives on the physical world.
For example, if we hear: \emph{joyful laughing, a playful scream, and a splash;}
we not only can visualize 
literal objects / actions that might have given rise to the audio scene, 
but also, we can reason about plausible higher-level facets, e.g.,
a child speeding down a water slide at a water park, splashing through the water (see Figure~\ref{fig:image_pivot_idea}).

Machines capable of parsing, representing, and describing such environmental sound hold practical promise. For example, according to the National Association of the Deaf's captioning guide, accessible audio caption generation systems should go beyond speech recognition (i.e., identifying speakers and transcribing the literal content of their speech) and 
provide the textual description of all the sound effects, e.g., ``a large group of people talking excitedly at a party'', in order to provide the full information contained in that audio.\footnote{{\href{https://www.nad.org/resources/technology/captioning-for-access/what-is-captioning/}{nad.org's captioning guide}}; \citet{gernsbacher2015video} discusses the benefits of video captions beyond d/Deaf users.}

The dominant paradigm for studying \emph{machine hearing}~\citep{lyon2010machine} has been through human-annotated audio-text data, where text is either free-form audio descriptions (e.g., ``the sound of heavy rain'') or tagsets~\citep{US8K,audioset,audiocaps,audiocaps-clotho}.
But existing supervised audio-text resources are limited.
While some audio-text co-occurences can be sourced from audio-tag co-occurrences~\citep{freesound} or from video captioning data~\citep{rohrbach2015dataset,MSR-VTT,oncescu2021queryd}, 
they are either not sufficiently related to environmental sound or limited in their scale and coverage. 

\begin{figure}[t!]
  \centering\small
    \includegraphics[width=0.95\columnwidth]{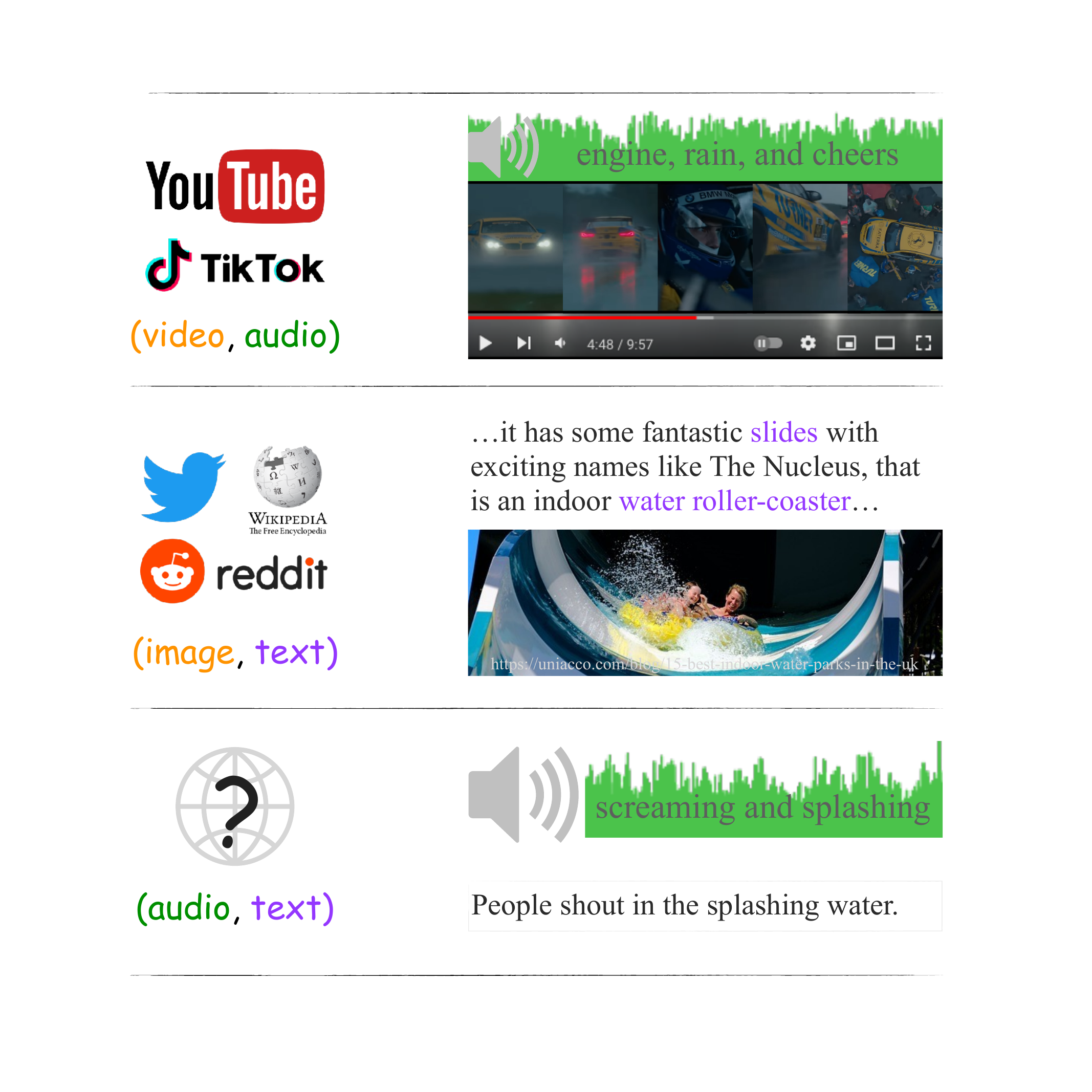}
\caption{\label{fig:keyidea}
Video-audio and image-text co-occurrences are abundantly available on the web to support the learning of video-audio alignment and image-text alignment (e.g., via large-scale video-audio and image-text pre-training), but audio-text co-occurrences are not.
}
\vspace*{-.25cm}
\end{figure}

In this paper, we study large-scale audio-text alignment without paired audio-text (AT) data. 
Inspired by pivot-based models for unsupervised machine translation~\citep{wu2007pivot,pivot-comparison-mt}, we propose \modelname, short for \modelnamelong. \modelname uses images as a pivot modality to connect audio and text.
It parallels our motivating example: hearing a sound, humans can visually \emph{imagine} the associated situation and literally \emph{describe} it.
Pivoting is practically viable because there are abundantly available image-text (VT) and video-audio (VA) co-occurrences on the web, from which bimodal correspondence models can be trained (see Figure~\ref{fig:keyidea}). By linking audio and text implicitly via the combination of the VT and VA models, we enable \emph{zero-resource} connection between audio and text, i.e., \modelname can reason about audio-text connections despite never having observed these modalities co-occur explicitly.



We evaluate on zero-shot audio-text retrieval and zero-shot audio classification. On the Clotho caption retrieval task~\citep{audiocaps-clotho}, without any parallel AT data, \modelname surpasses the supervised state of the art by 2.2\% R@1;
on zero-shot audio classification tasks, it establishes new state of the arts, achieving 57.1\% accuracy on ESC50~\citep{ESC50} and 44.7\% accuracy on US8K~\citep{US8K}.
We also show that the zero-resource pivoting AT model \modelname can be improved by:
\begin{enumerate}[wide=0\parindent,noitemsep]
\item[(1)] \emph{Unsupervised curation:} whereby noisy AT pairs are explicitly mined from the pivoting model and serve as additional training data (e.g., +5.7\% on ESC50 and +9.3\% on US8K); and
\item[(2)] \emph{Few-shot curation:} whereby a small number of human-annotated audio caption pairs are made available at training time (e.g., a few hundred pairs increases the zero-shot audio classification accuracy by 8\% on US8K).
\end{enumerate}

%


However, for ESC-50, according to the empirical scaling relationship we find, it would require around $2^{21} \approx 2\text{M}$ aligned audio-text pairs for the zero-shot model to match human parity on ESC50 under our setup, which is an order-of-magnitude more than the largest currently-available audio-text corpus of~\citet{audiocaps}.

\section{Related work}

\newcommand{\xmark}{\ding{55}}%
\begin{table*}[t]\small
\centering
{\setlength{\tabcolsep}{0.2em} 
\makebox[\linewidth]{\resizebox{1.\linewidth}{!}{%
\begin{tabular}{
	@{} 
	m{0.001\linewidth} @{\hspace{0.0cm}} 
	p{0.27\linewidth} @{\hspace{0.0cm}} 
	M{0.18\linewidth}  @{\hspace{0.0cm}} 
	M{0.1\linewidth} @{\hspace{0.0cm}} 
	M{0.15\linewidth} @{\hspace{0.0cm}} 
	M{0.1\linewidth} @{\hspace{0.1cm}} 
	M{0.1\linewidth} @{\hspace{0.0cm}} 
	m{0.001\linewidth} 
	@{}
}
& Model & AE Initialization & Objective & AT Supervision  & VT Alignment & Zero-shot AT Retrieval \\
\toprule
& MMV~\citep{MMV} & Random & $\gL_{bi\text{-}bi}$ &None &  Trainable &  \xmark \\
& VATT~\citep{VATT} & Random &  $\gL_{bi\text{-}bi}$ & None & Trainable &  \xmark\\
& AudioCLIP~\citep{audioclip} & ImageNet
&  $\gL_{tri}$ & 2M Audio Tags & Trainable &  \xmark\\
& Wav2CLIP~\citep{wav2clip} & Random & $\gL_{bi\text{-}bi}$ & None  &  Frozen &  \xmark \\
& \modelname (ours) & Image CLIP & $\gL_{bi\text{-}bi}$ & None &  Frozen & \checkmark \\
& \modelname+AT (ours) & Image CLIP & $\gL_{bi\text{-}bi}$ & {Caption Curation} &  Frozen & \checkmark \\
\bottomrule
\end{tabular}}}}								
\caption{\label{tab:model_comparison}
Survey of recent prior work studying for tri-modal (images, audio, and text) representation learning. AE is short for \textbf{A}udio \textbf{E}ncoder. Some work experiments with more than one objective, we report the best or the one it advocates. Importantly, we report zero-shot audio-text retrieval between audio and full-sentence text descriptions, along with scaling laws associated with that setup.
}
\vspace*{-.0cm}
\end{table*}

\paragraph{Supervised audio representation learning.} While automatic speech recognition has been a core focus of the audio processing community,
environment sound classification has emerged as a new challenge and is drawing more attention~\citep{US8K,ESC50,audioset}.
Some prior work in learning sound event representations are supervised by category labels
~\citep{DeepCNN4Waveform,SharedImageAudioEncoder,CNN4Audio,ResNet4Audio,AST}. Others use weaker forms of supervision for tagging~\cite{kumar2017audio,kong2018audio} and localization~\cite{mcfee2018adaptive,kim2019sound}. 

\paragraph{Learning audio representations from visual imagination.}
There are two main paradigms for using visual information to derive audio representations.
In the two-stage setup, an image encoder is first pre-trained; 
these weights are used as the initialization of the supervised audio model~\citep{ResNet4Audio,AST}. 
The other adopts contrastive learning: it exploits the image-audio alignment inherent in videos and learns audio and image / video representations jointly~\citep{korbar2018cooperative,wang2021multimodal,MBT}.
We use insights from both directions by (1) using CLIP's image encoder, which has been pre-trained on image-text pairs~\citep{CLIP}, to initialize an audio encoder and (2) using contrastive pre-training on image-audio pairs.
Throughout training, we do not require any labeled images or audio.

\paragraph{Tri-modal learning of audio-text alignment.} Our work extends recent work that generalizes the bi-modal contrastive learning  to a tri-modal setting~\citep{MMV,VATT}.
While they also 
connect audio and text implicitly by using images as a pivot, the quality of this 
audio-text alignment has rarely been studied.
To our knowledge, we present the first comprehensive evaluation of the inferred audio-text alignment via zero-shot retrieval / classification.

The work closest to ours are AudioCLIP~\citep{audioclip} and Wav2CLIP \cite{wav2clip}. AudioCLIP's pre-training setup is similar to ours, but requires human-annotated textual labels of audio, while ours does not. Wav2CLIP is concurrent with our work; while similar-in-spirit,
our model not only performs significantly better, but also, we more closely explore methods for improving audio-text alignment, e.g., unsupervised curation.

\paragraph{Pivot-based alignment models.}
The pivoting idea for alignment learning can
date back to~\citet{pivot_idea}.
Language pivots~\citep{wu2007pivot,pivot-comparison-mt} and image pivots~\citep{specia2016shared,hitschler2016multimodal,nakayama2017zero} have been explored in zero-resource machine translation.
Pivot-based models have also been shown to be helpful in learning image-text alignment~\cite{li2020oscar}.

\section{Model}
We first formalize tri-modal learning by assuming available co-occurrence data for every pair of modalities (\S~\ref{sec:model-tri-modal}). Then we present bi-bi-modal pre-training as an alternative when there is no paired audio-text data, and
implement \modelname via bi-bi-modal pre-training (\S~\ref{sec:model-vip-ant}). Finally, we describe model variants for cases of varying AT supervision (\S~\ref{sec:model-audio_text_aug}). 

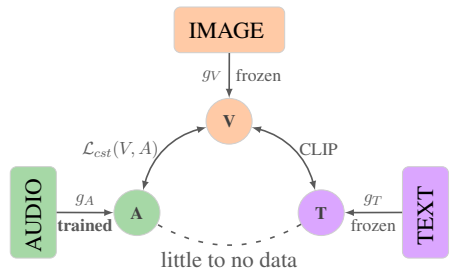
\begin{figure}
\begin{center}
\resizebox{0.75\linewidth}{!}{%
\begin{tikzpicture}[node distance=0mm]

\node[circle, fill=mygreen!35,draw=lightgray!75,anchor=north west,minimum width=1.cm,minimum height=1.cm] (A) at (2, 4) {\textcolor{black!75}{\bf A}};

\node[vmodule,fill=mygreen!35,minimum width=20mm] (AA) at ([shift=({-2.25cm,-1cm})]A.west) { AUDIO };

\node[circle, fill=myorange!35,draw=lightgray!75,anchor=north west,minimum width=1.cm,minimum height=1.cm] (V) at (4, 6) {\textcolor{black!75}{\bf V}};

\node[hmodule,fill=myorange!35,minimum width=24mm] (VV) at ([shift=({-1.2cm,2cm})]V.north) { IMAGE };

\node[circle, fill=mypurple!35,draw=lightgray!75,anchor=north west,minimum width=1.cm,minimum height=1.cm] (L) at (6, 4) {\textcolor{black!75}{\bf T}};

\node[vmodule,fill=mypurple!35,minimum width=20mm] (LL) at ([shift=({1.25cm,-1cm})]L.east) { TEXT };

\draw [>=latex,<->,line width=.35mm,black!65] (A) to [bend left] node[left]{$\gL_{cst}(V, A)$} (V);
\draw [>=latex,<->,line width=.35mm,black!65] (V) to [bend left] node[right]{CLIP} (L);
\draw [loosely dashed,line width=.35mm,black!65] (L) to [bend left] node[below,font=\fontsize{14pt}{16pt}\selectfont]{little to no data} (A);

\draw [-latex,line width=.35mm,black!65] (AA) to node[above]{$g_A$} node[below]{\textcolor{black!65}{\bf trained}}  (A);
\draw [-latex,line width=.35mm,black!65] (LL) to node[above]{$g_T$} node[below]{frozen}  (L);
\draw [-latex,line width=.35mm,black!65] (VV) to node[left]{$g_V$} node[right]{frozen}  (V);
\end{tikzpicture}}
\end{center}
\caption{\label{fig:bibi-modal_diagram}
Learning paradigm of \modelname.
}
\vspace*{-.0cm}
\end{figure}

\subsection{Tri-modal representation learning}\label{sec:model-tri-modal}
Tri-modal representation learning between images, audio, and text aims to derive representations from co-occurrence patterns among the three modalities~\citep{MMV,VATT}.
We consider a simple tri-modal representation space, 
which relies on encoding functions $g_V: V\rightarrow\rmV$, $g_A: A\rightarrow\rmA$, and $g_T: T\rightarrow\rmT$ to map images $v$, audio $a$, and text $t$ ($v\in V, a\in A,\text{ and } t\in T$), respectively, to a shared vector space: $\vv, \va, \vt\in\sR^{d}$ ($\vv\in\rmV, \va\in\rmA,\text{ and } \vt\in\rmT$).
Instead of pre-specifying the precise semantics of this 
continuous space, vector similarities across modalities are optimized to reconstruct co-occurrence patterns in training corpora, i.e., two vectors should have a higher dot product if they are more likely to co-occur.
We use contrastive learning with the InfoNCE loss \cite{sohn2016improved,oord2018representation}:
\begin{flalign}\label{eq:cst}
&\gL_{cst}(A, B)  = \nonumber \\ \!\!\!\sum_{i}
&\frac{\exp{ s(\va^{(i)}, \vb^{(i)})}}{ \sum_{\va}\exp{s(\va, \vb^{(i)})}} +
\frac{\exp{ s(\va^{(i)}, \vb^{(i)})}}{ \sum_{\vb}\exp{s(\va^{(i)}, \vb)}} \,,
\end{flalign}
where $A, B$ are two sets of data points from two different modal domains, respectively; $\va^{(i)}, \vb^{(i)}$ are vector representations of the co-occuring pair $(a^{(i)}, b^{(i)})$ which are encoded by $g_A(a^{(i)})$ and $g_B(b^{(i)})$, respectively; $s(\va, \vb)$ computes the similarity between $\va$ and $\vb$, which we take to be scaled cosine similarity.

If we had access to co-occurrence data between all pairs of modalities, we could optimize the tri-modal loss:
\begin{flalign}\label{eq:tri}
&\gL_{tri}(V, A, T) = \nonumber \\
&\gL_{cst}(V, A) + \gL_{cst}(A, T) + \gL_{cst}(V, T)\,.
\end{flalign}

\subsection{Visually pivoted audio and text}\label{sec:model-vip-ant}

Differently from image-text and image-audio pairs, which are abundantly available on the web, audio-text data is scarce.
Instead of Equation~\ref{eq:tri}, in \modelname, we consider a ``bi-bi-modal" loss, which doesn't require AT data.
\begin{flalign}\label{eq:bibi}
\gL_{bi\text{-}bi}(V, A, T) = \gL_{cst}(V, A) + \gL_{cst}(V, T)\,.
\end{flalign}

The image encoder is shared between the VA alignment model (i.e., $\gL_{cst}(V, A)$) and the VT alignment model (i.e., $\gL_{cst}(V, T)$) and thus provides a zero-resource connection between audio and text in the tri-modal embedding space implicitly.

\subsubsection{Model architecture}

\paragraph{Image and text encoders.} 
Instead of learning $g_V$ and $g_T$ from scratch, we build on a pre-trained CLIP model, which
has been pre-trained on WebImageText (WIT), a dataset of 400 million image-text pairs gathered from the internet~\citep{CLIP}. CLIP has been shown highly performant on VT tasks, e.g., zero-shot image classification. We use the ViT-B/32 model in this work, which consists of a 12-layer vision Transformer (ViT) and a 12-layer language Transformer \cite{Transformer,ViT}.
Given CLIP's strong VT alignment,
we use its image encoder as $g_V$ and text encoder as $g_T$.
During learning, $g_V$ and $g_T$ are kept frozen and thus the joint VT representation space is untouched (see Figure~\ref{fig:bibi-modal_diagram}).
We minimize only the first loss term of Equation~\ref{eq:bibi}:
\begin{flalign}\label{eq:bibi-ours}
\min_{\Theta_A} \gL_{cst}(V, A)\,,
\end{flalign}
where $\Theta_A$ are the trainable parameters of the audio encoder $g_A$.

\begin{figure}
	\begin{center}
		\resizebox{0.99\linewidth}{!}{%
			\includegraphics[width=0.98\linewidth]{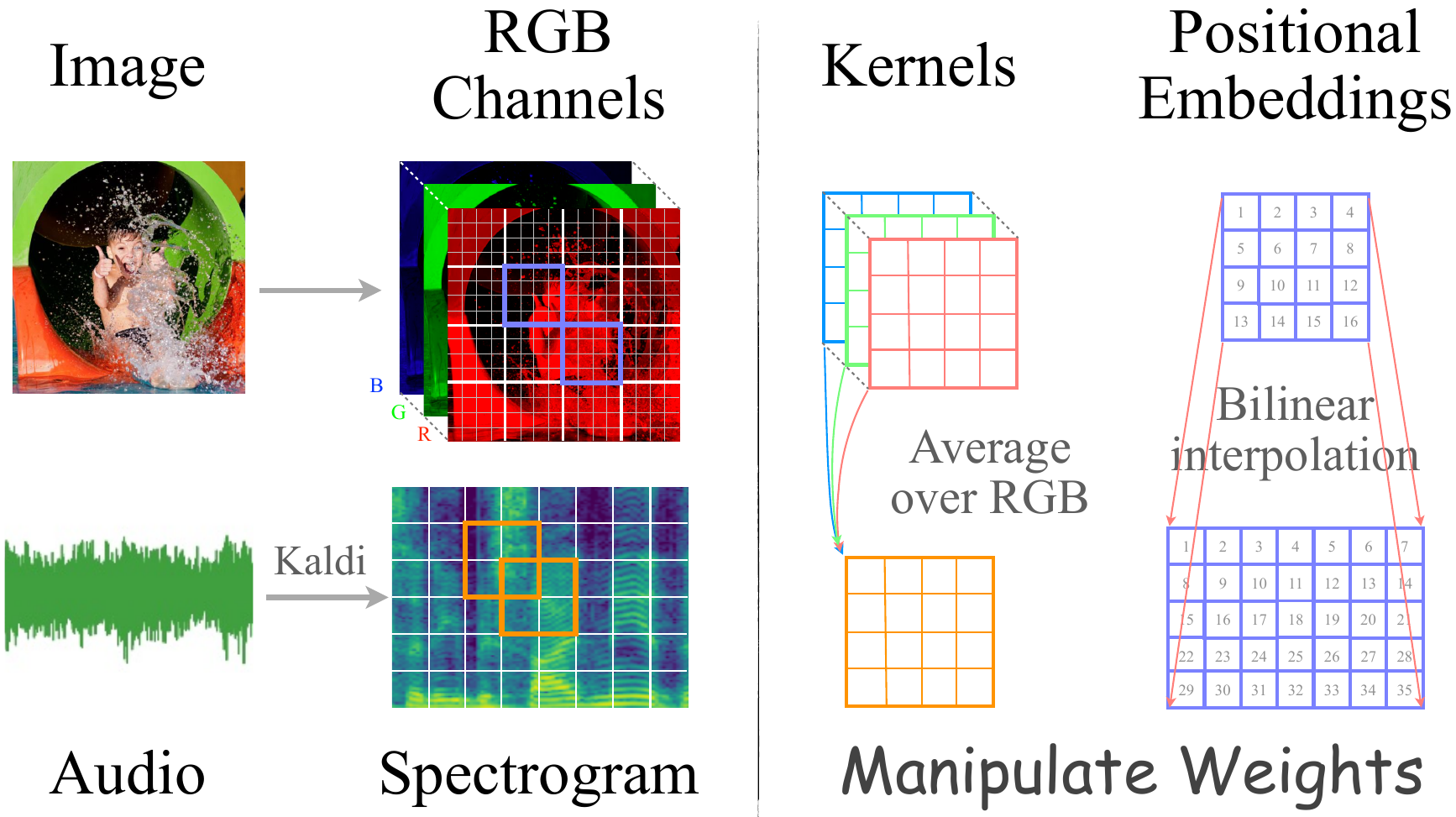}
		}
	\end{center}
	\caption{\label{fig:clip_init}
\textbf{Left:} three-channel image versus one-channel Spectrogram features of audio. We use ViT~\citep{ViT} to encode images and audio. ViT uses a convolution layer to encode non-overlapped image patches into a sequence of image tokens, but for audio we modify the convolution stride to allow for overlaps between neighbor patches. \\
\textbf{Right:} adapting the convolution layer of ViT for audio encoding. 
For simplicity's sake, we omit the output channels of kernel weights and positional embeddings. 
}
\vspace*{-.0cm}
\end{figure}

\paragraph{Audio encoder.} 
Our audio encoder has the same vision Transformer architecture as CLIP's image encoder (\text{ViT-B/32}). In \S~\ref{sec:exp}, we show that initializing the audio encoder with CLIP's visual weights significantly improves convergence speed and accuracy.
The architectural modifications which enable the use of visual CLIP's architecture for audio are (see Figure~\ref{fig:clip_init} for an illustration):\footnote{\href{ https://github.com/zhaoyanpeng/vipant}{https://github.com/zhaoyanpeng/vipant}}
\begin{itemize}[wide=0\parindent,noitemsep]
\item[(1)] We customize the convolution stride to allow for overlaps between neighbor patches of Spectrogram features of audio.
\item[(2)] In the input embedding layer, we average the kernel weights of the convolution layer along the input channel to account for 1-channel Mel-filter bank features of audio (\emph{cf}. RGB channels of images).
\item[(3)] We up-sample the 2-dimensional positional embeddings of image tokens to account for longer audio token sequences.
\end{itemize}


\subsubsection{Bi-bi-modal pre-training details}\label{app:bi-bi}

\begin{figure}[!t]
	\begin{center}
		\includegraphics[width=0.98\linewidth]{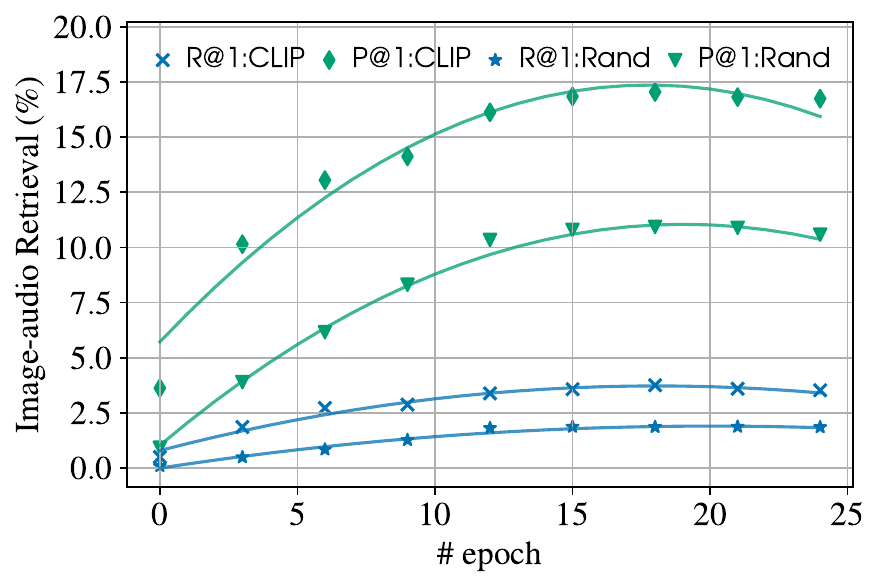}
	\end{center}\vspace{-.25em}
	\caption{\label{fig:pnr_pretrain_va}
		Image $\rightarrow$ Audio retrieval performance per image-audio pre-training epoch, evaluated on the AS balanced training set. "CLIP" and "Rand" indicates that the audio encoder is initialized from CLIP's image encoder and has random initialization, respectively.
		}
		\vspace{-.0em}
\end{figure}

\paragraph{Video-audio co-occurences.} 
To optimize Equation~\ref{eq:bibi-ours}, we gather VA co-occurrences from AudioSet (AS;~\citet{audioset}),\footnote{\href{https://github.com/zhaoyanpeng/audioset-dl}{https://github.com/zhaoyanpeng/audioset-dl}}
which contains temporally aligned audio and video frames from 10-second clips gathered from around 2 million YouTube videos.
To construct aligned image-audio pairs from AS, we adopt a sparse sampling approach \cite{lei2021less}: we first, extract four equal-spaced video frames from each clip. 
Then, during training, we randomly sample a frame from the four, and treat it as co-occurring with the corresponding audio clip.
At test time, we always use the second video frame as the middle frame to construct image-audio pairs.
We use the unbalanced training set, which consists of around 2 million video clips, to pre-train the audio encoder.
Since AudioSet does not provide an official validation set, we validate the audio encoder and tune model hyperparameters on the balanced training set.

\paragraph{Audio preprocessing.}
We use Kaldi~\citep{kaldi} to create Mel-filter bank features (FBANK) from the raw audio signals. Specifically, we use the Hanning window, 128 triangular Mel-frequency bins, and 10 millisecond frameshift. We always use the first audio channel when an audio clip has more than one channel.
We apply two normalizations: (1) before applying Kaldi, we subtract the mean from the raw audio signals; and (2) we compute the mean and standard deviation of FBANK on the unbalanced AS training set, and then normalize the FBANK of each audio clip.
For data augmentation, inspired by~\citet{AST},
we use frequency masking and time masking:
we randomly mask out one-fifth FBANK along the time dimension and one-forth FBANK along the frequency dimension during training.

\paragraph{Training dynamics.}
The architecture of our audio encoder follows the vision Transformer of CLIP (ViT-B/32, see \citet{CLIP} for more details).
For the trade-off of efficiency and efficacy, we set the convolution stride to $16\times 24$.
This results in around 300 audio tokens for a kernel size of $32\times 32$ and an input size of $1000\times 128$ (all in the form of $\textit{time}\times\textit{frequency}$).
We optimize the model with LARS~\citep{LARS},
where the initial learning rates for model weights and model biases are set to 2e-1 and 4.8e-3, respectively (detailed hyperparameters can be found in Table~\ref{tab:optimizer_setups} in Appendix~\ref{app:hyperparam}).
We pre-train our model on 4 NVIDIA Quadro RTX 8000 GPUs and for 25 epochs.
We empirically set the batch size to 432 to fit the GPU memory.
The full pre-training can be done within 24 hours.

\paragraph{Evaluation.} 
We measure the VA pre-training performance by retrieval precision and recall:
\begin{align}
    p &= \frac{\#\text{(relevant items among the retrieved)}}{\#\text{(retrieved items)}}\,,\nonumber \\
    r &= \frac{\#\text{(relevant items among the retrieved)}}{\#\text{(relevant items)}}\,.\nonumber
\end{align}
Audio is relevant if it has the same set\footnote{Recall that each audio clip in AudioSet is annotated with multiple labels.} of labels as the image query, and vice versa.
We average precisions and recalls over all samples in the balanced AS training set.
Figure~\ref{fig:pnr_pretrain_va} illustrates the top-1 retrieval performance with images as the query (similar trends are observed when using audio as the query). 
Compared with random initialization, initializing the audio encoder from CLIP's image encoder leads to faster convergence and better VA alignment.
As we will see, this performance on VA retrieval transfers to downstream AT tasks. 

\subsection{Unsupervised and few-shot curation}\label{sec:model-audio_text_aug}
\begin{table*}[t]\small
	\centering
	{\setlength{\tabcolsep}{0pt} 
		\makebox[\linewidth]{\resizebox{1.\linewidth}{!}{%
				\begin{tabular}{@{} m{0.03\linewidth} @{\hspace{0.0cm}} m{0.08\linewidth} @{\hspace{0.0cm}} m{0.75\linewidth} @{}}
					\toprule
					
					\multirow{12}{*}{\rotatebox[origin=c]{90}{{\footnotesize Unsupervised (Zero-resource)}}}
					&\textbf{AC}  & \textbf{A}udio-focused \textbf{C}aptions originate from the training captions of AudioCaps and Clotho. We perform caption retrieval by using CLIP and the prompt "the sound of". {(1080078 aligned pairs)} \\

					&example & \coloredrow{1}{olive!5}{\textit{A balloon is rubbed quickly and slowly to make squeaking sounds.}} \\
					\cmidrule{2-3}
					
					&\textbf{FC} & \textbf{F}ree \textbf{C}aptions are generated by priming GPT-J with MSCOCO captions. We perform caption retrieval by using CLIP and the prompt "a photo of". {(1224621 aligned pairs)} \\

					&example & \coloredrow{1}{olive!5}{\textit{The blue colored person is jumping on the white and yellow beach ball.}} \\					
					\cmidrule{2-3}
					&\textbf{VC} & \textbf{V}ision-focused \textbf{C}aptions originate from MSCOCO. We perform caption retrieval by using CLIP and the prompt "a photo of". {(1172276 aligned pairs)}\\

					&example & \coloredrow{1}{olive!5}{\textit{A sky view looking at a large parachute in the sky.}}\\
					\cmidrule{2-3}
					&\textbf{RC} & \textbf{R}andom \textbf{C}aptions indicates that we break the gold AL alignment in AudioCaps by randomly sampling a caption for each audio clip. They are used as a lower bound on the quality of AL alignment. {(44118 aligned pairs)} \\
					&example & \coloredrow{1}{olive!5}{\textit{A whoosh sound is heard loudly as a car revs its engines.}} \\

					\cdashlinelr{1-3}

					\multirow{5}{*}{\rotatebox[origin=c]{90}{{\footnotesize Supervised}}}
					&\textbf{GL} & \textbf{G}old textual \textbf{L}abels are used to construct AL pairs. {(120816 aligned pairs)} \\

					&example & \coloredrow{1}{olive!5}{\textit{Gurgling}} \\
					\cmidrule{2-3}
					&\textbf{GC} & \textbf{G}old \textbf{C}aptions from AudioCaps provide an upper bound on the quality of AL alignment. {(44118 aligned pairs)} \\

					&example &\coloredrow{1}{olive!5}{\textit{Children screaming in the background as the sound of water flowing by.}} \\
					\bottomrule
	\end{tabular}}}}
	\centering
	{\setlength{\tabcolsep}{0pt} 
	\makebox[\linewidth]{\resizebox{1.\linewidth}{!}{%
			\begin{tabular}{@{} 
					m{0.097\linewidth} @{\hspace{0.003\linewidth}} 
					m{0.097\linewidth} @{\hspace{0.003\linewidth}} 
					m{0.097\linewidth} @{\hspace{0.003\linewidth}} 
					m{0.097\linewidth} @{\hspace{0.003\linewidth}} 
					m{0.097\linewidth} @{\hspace{0.003\linewidth}} 
					m{0.097\linewidth} @{\hspace{0.003\linewidth}} 
					m{0.097\linewidth} @{\hspace{0.003\linewidth}} 
					m{0.097\linewidth} @{\hspace{0.003\linewidth}} 
					m{0.097\linewidth} @{\hspace{0.003\linewidth}} 
					m{0.097\linewidth} 
					@{}}
	\raisebox{-\totalheight}{\includegraphics[width=\linewidth]{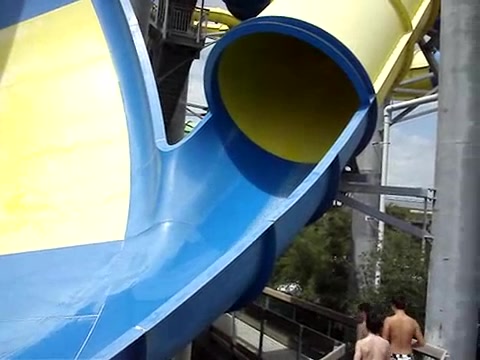}} &
	\raisebox{-\totalheight}{\includegraphics[width=\linewidth]{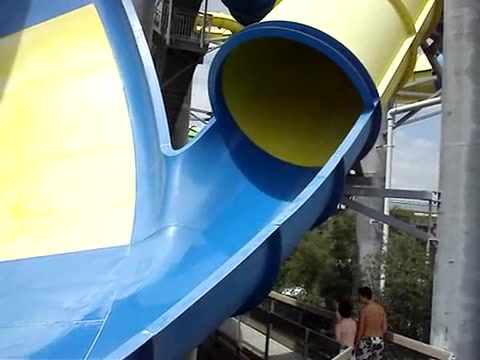}} &
	\raisebox{-\totalheight}{\includegraphics[width=\linewidth]{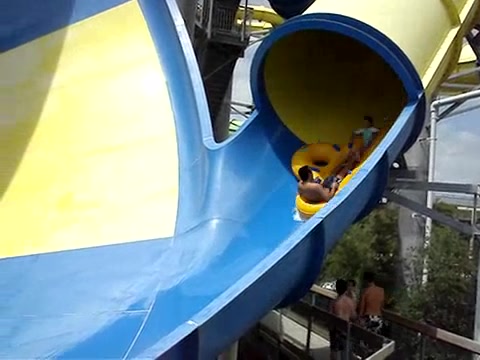}} &
	\raisebox{-\totalheight}{\includegraphics[width=\linewidth]{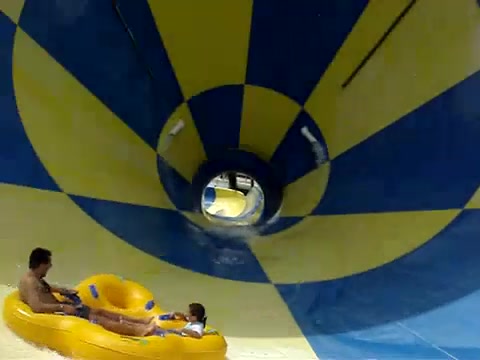}} & \raisebox{-\totalheight}{\includegraphics[width=\linewidth]{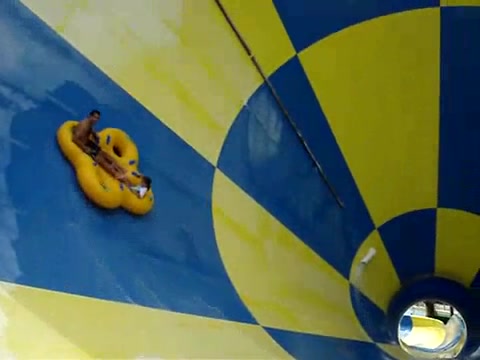}} &
	\raisebox{-\totalheight}{\includegraphics[width=\linewidth]{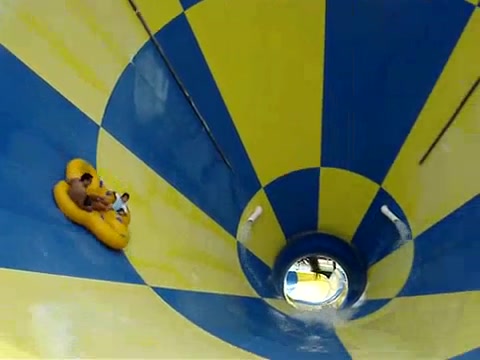}} &
	\raisebox{-\totalheight}{\includegraphics[width=\linewidth]{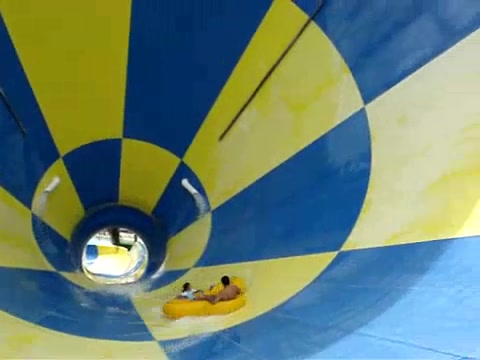}} & \raisebox{-\totalheight}{\includegraphics[width=\linewidth]{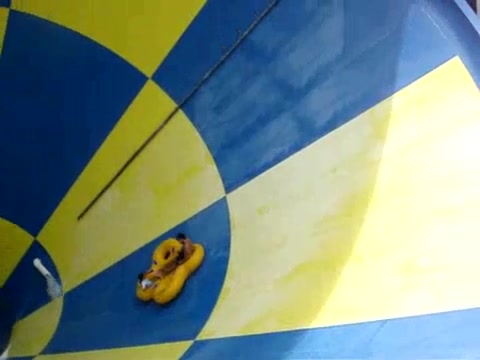}} &
	 \raisebox{-\totalheight}{\includegraphics[width=\linewidth]{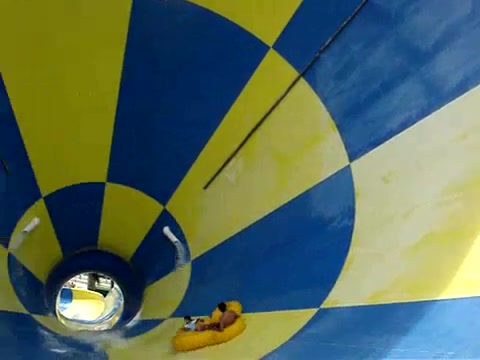}} &
	 \raisebox{-\totalheight}{\includegraphics[width=\linewidth]{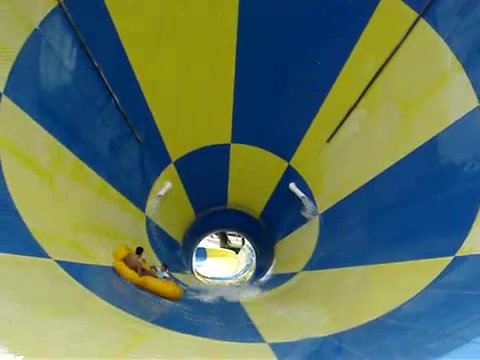}} \\
	\end{tabular}}}}								
	\caption{\label{tab:al_curation}
		Different ways of curating AT pairs. \textit{Gurgling} is described as "the bubbling sound of water flowing through a narrow constriction, such as from a bottle with a narrow neck". The example comes from this YouTube video: \href{https://www.youtube.com/watch?v=1O7-QuhweZE}{1O7-QuhweZE}.
	}
	\vspace{-.0em}
\end{table*}

To improve the AT alignment beyond pivoting, we consider curating
audio-text pairs, and then performing an additional fine-tuning step by training the audio encoder with the AT loss, i.e., $\gL_{cst}(A, T)$.\footnote{Since our goal is to improve AT alignment, we primarily focus on AT fine-tuning; nonetheless, we compare AT fine-tuning to full VAT fine-tuning as in Equation~\ref{eq:tri} in Appendix~\ref{app:val_versus_al_finetuning}.} During AT fine-tuning, we keep the text encoder $g_T$ frozen and only fine-tune the audio encoder. 

\paragraph{Unsupervised curation.} We consider explicitly mining AT pairs from \modelname. Because this zero-resource method uses no human supervision, we refer to it as ``unsupervised curation." Concretely, for each video segment in AudioSet, we extract a video frame, and input that frame to the original CLIP image encoder. Then, we encode a large set of candidate captions, and perform Image $\rightarrow$ Text retrieval over them by using the CLIP text encoder. The top candidate captions according to cosine similarity are then paired with the audio that corresponds to the original video clip.

We consider multiple caption sources to search over. As noted by \newcite{audiocaps}, captions for images and captions for environmental audio are significantly different in focus. We consider two vision-focused caption sets: (1) MSCOCO~\citep{MSCOCO} captions (VC); and (2) because MSCOCO captions are limited to 80 object categories, we generate free-captions from GPT-J \citep{gpt-j} conditioned on MSCOCO captions as a prompt (FC). We additionally consider audio-focused captions from the training set of AudioCaps~\citep{audiocaps} and Clotho \cite{audiocaps-clotho} (AC).\footnote{We do not use the \emph{alignment} of these captions --- just the captions themselves.} As a baseline, we also consider a random caption alignment, which assigns a random caption from AC to each clip (instead of pivoting on images). The upper half of Table~\ref{tab:al_curation} summarizes different ways of curating AT pairs without additional supervision.

\paragraph{Few-shot curation.}
We also explore the effect of incorporating limited amounts of AT supervision, specifically, 
via captions from AudioCaps (GC) and textual labels of AudioCaps (GL) (see the bottom half of Table~\ref{tab:al_curation}).

\section{Audio-text experiments}\label{sec:exp}

\begin{table*}[!t]\small
	\centering
	{\setlength{\tabcolsep}{1.2em} 
		\makebox[\linewidth]{\resizebox{1.\linewidth}{!}{%
				\begin{tabular}{p{0.01cm} p{2.5cm} rrrrrrrr}
					\toprule
					& \multirow{3}{*}{Model} & \multicolumn{4}{c}{AudioCaps} & \multicolumn{4}{c}{Clotho} \\
					\cmidrule{3-10}
					& & \multicolumn{2}{c}{Text$\rightarrow$Audio} & \multicolumn{2}{c}{Audio$\rightarrow$Text} &  \multicolumn{2}{c}{Text$\rightarrow$Audio} & \multicolumn{2}{c}{Audio$\rightarrow$Text}\\ 
					& & {\scriptsize R$@$1} & {\scriptsize R$@$10} & 
					{\scriptsize R$@$1} & {\scriptsize R$@$10} &   
					{\scriptsize R$@$1} & {\scriptsize R$@$10} & 
					{\scriptsize R$@$1} & {\scriptsize R$@$10}\\ 
					\midrule
					& Supervised SoTA & $18.0$&$62.0$&$ 21.0$&$62.7$ & $4.0$&$25.4$&$ 4.8 $&$25.8$ \\
					\midrule
					\multirow{6}{*}{\rotatebox[origin=c]{90}{{\footnotesize Zero-resource}}} & VA-Rand & 1.3 & 7.3 & 5.6 & 24.5 & 1.3 & 7.5 & 3.2 & 13.5 \\
					& \modelname & 0.8 & 7.9 & 10.1 & 38.1 & 1.9 & 9.5 & 7.0 & 25.6\\
					&  +AT w/ AC     & 9.9 & 45.6 & 15.2 & 52.9  & 6.7 & 29.1 & 7.1 & 30.7\\
					&  +AT w/ FC     & 8.9 & 41.5 & 14.7 & 50.0   & 6.5 & 27.7 & 7.8 & 29.7 \\
					&  +AT w/ VC     & 6.9 & 35.7 & 13.5 & 49.4 & 5.5 & 25.6 & 7.6 & 28.2 \\
					&  +AT w/ RC     & 3.8 & 19.9 & 10.7 & 38.1 & 3.5 & 16.9 & 5.5 & 24.9 \\

					\cdashlinelr{2-10}
					\multirow{2}{*}{\rotatebox[origin=r]{90}{\parbox{.3cm}{Zero-shot}}}
					&  +AT w/ GL     & 12.4 & 52.9 & 13.0 & 51.2 & 6.7 & 29.0 & 6.8 & 27.0 \\ 
					&  +AT w/ GC     & \textbf{27.7} & \textbf{78.0} & \textbf{34.3} & \textbf{79.7} & \textbf{11.1} & \textbf{40.5} & \textbf{11.8} & \textbf{41.0}  \\

					\midrule
					& OracleAV-CLIP & 4.8 & 27.8 & 6.6 & 31.2 &\phantom{.}&\phantom{.}&\phantom{.}&\phantom{.}\\
					\bottomrule
	\end{tabular}}}}
	\caption{\label{tab:audiocap-retrieval_test}
		Audio caption retrieval performance (\%) on AudioCaps test set and Clotho evaluation set. "Supervised SoTA" is from~\citet{audio_retrieval}. OracleAV-CLIP: we replace audio with the corresponding image and evaluate image-text retrieval performance of CLIP~\citep{CLIP}.
		VA-Rand is the same as \modelname but he audio encoder is initialized randomly, instead of from CLIP visual weights. We further fine-tune \modelname on AT data curated in different fashions, e.g., AC, FC, VC, and RC are mined explicitly from the zero-resource pivoting model (see Table~\ref{tab:al_curation} for details).
	}
	\vspace{-.0em}
\end{table*}
\begin{table}[!t]
	{\setlength{\tabcolsep}{.2em}
		\makebox[\linewidth]{\resizebox{\linewidth}{!}{%
				\begin{tabular}{
						@{} p{0.4cm}
						@{\hspace{0.05cm}} p{2.7cm} 
						@{\hspace{0.35cm}} p{1.65cm} 
						@{\hspace{0.25cm}} p{1.65cm} 
						@{\hspace{0.25cm}} p{1.55cm} 
						@{}
					}

					&  Model & ESC50 & US8K & AS \\ \midrule
					&  Supervised & \text{95.7}$_{\pm 1.4}$ & \text{86.0}$_{\pm 2.8}$ & \text{37.9} \\
                    \midrule
					\multirow{7}{*}{\rotatebox[origin=c]{90}{{\footnotesize Zero-resource}}} &
					
					VA-Rand & 37.6{\footnotesize (33.0)} & 41.9{\footnotesize (38.1)} & \leavevmode{\phantom{0}}1.7{\footnotesize (\phantom{0}2.0)} \\
					&  \modelname     &   57.1{\footnotesize (49.9)}   & 44.7{\footnotesize (37.8)}  &  \leavevmode{\phantom{0}}2.6{\footnotesize (\phantom{0}2.8)} \\
                    
                    &  +AT w/ AC      &   62.8{\footnotesize (55.7)}   &  54.0{\footnotesize (47.0)}  &  11.6{\footnotesize (12.3)} \\
	 		 		&  +AT w/ FC     &  62.5{\footnotesize (58.0)}    & 52.7{\footnotesize (50.0)}   & 11.2{\footnotesize (12.2)} \\
	 		 		&  +AT w/ VC     &   61.9{\footnotesize (58.0)}   &  52.7{\footnotesize (50.3)}  &  \leavevmode{\phantom{0}}8.9{\footnotesize (10.7)} \\
	 		 		&  +AT w/ RC     &   51.6{\footnotesize (36.1)}    &  42.3{\footnotesize (28.5)}   &  \leavevmode{\phantom{0}}4.1{\footnotesize (\phantom{0}4.6)} \\
	 		 		& Wav2CLIP & 41.4 & 40.4 &  \\

					\cdashlinelr{2-5}
                    \multirow{3}{*}{\rotatebox[origin=c]{90}{{\footnotesize Zero-shot}}}
	 		 		&  +AT w/ GL     &   67.2{\footnotesize (64.5)}    &  62.6{\footnotesize (61.0)}   &  15.4{\footnotesize (\textbf{18.9})}  \\
					&  +AT w/ GC     &  \textbf{69.5}{\footnotesize (64.2)}    &  \textbf{71.9}{\footnotesize (67.1)}  & 13.3{\footnotesize (13.6)} \\

					& AudioCLIP & \text{69.4} & 65.3 &  \\
					\bottomrule
	\end{tabular}}}}
	\caption{\label{tab:zeroshot-clf-merged-prompt}
		Zero-shot audio classification accuracies (\%) on ESC50 and US8K and mAPs (\%) on AudioSet (AS). 
		"Supervised" = upper bound performance of \modelname when fine-tuned with supervised audio labels.
		In the zero-shot / zero-resource settings, we use a prompt `\textit{the sound of}' by default (results in parenthesis are without the prompt). "+AT" = fine-tuned \modelname on AT pairs with different curations.
		AudioCLIP is pre-trainined using the 2 million textual labels of AudioSet; +AT w/ GL and +AT w/ GC are trained with only 44K labels / captions. Wav2CLIP is most directly comparable to our zero-resource pivoting model \modelname with unsupervised curation.
	}
	\vspace{-.0em}
\end{table}

We use two types of tasks to evaluate the quality of the audio-text alignments learned by our model: AT retrieval and zero-shot audio classification.

\paragraph{AT retrieval.} 
We conduct audio-text retrieval on AudioCaps and Clotho for in-domain evaluation and out-of-domain evaluation, respectively:
\begin{itemize}[wide=0\parindent,noitemsep]
\item[(1)] \textbf{AudioCaps}~\citep{audiocaps} builds on AudioSet~\citep{audioset} and provides captions for a subset of audio clips in AudioSet (sourced from YouTube). As we have pre-trained the audio encoder on AudioSet, we consider audio-text retrieval on AudioCaps as \emph{in-domain} evaluation.
\item[(2)] \textbf{Clotho}~\citep{audiocaps-clotho} consists of audio clips which have a duration of 15-30 seconds and come from Freesound~\citep{freesound}. It has a different sound source from AudioCaps and is used for \emph{out-of-domain} evaluation.
\end{itemize}

We study the out-of-domain generalizability of our models by applying them to Clotho directly, without further fine-tuning on it.\footnote{Clotho audio clips (15-30s) are longer than our pre-training audio clips (10s). See Appendix~\ref{app:interp_pos} for adaptation details.}

\paragraph{Zero-shot audio classification.}
\label{sec:audio-clf}
\label{sec:clf_datasets}
We consider the following three widely used datasets for audio classification.
\begin{itemize}[wide=0\parindent,noitemsep]
\item[(1)] \textbf{ESC50}~\citep{ESC50} contains 2000 audio clips from 50 classes.
Each audio clip has a duration of 5 seconds and a single textual label. We follow the standard $k$-fold data splits.
\item[(2)] \textbf{US8K}~\citep{US8K} contains 8732 audio clips from 10 classes.
Each audio clip has a duration less than 4 seconds and a single textual label. We follow the standard $k$-fold data splits.
\item[(3)] \textbf{AudioSet}~\citep{audioset} is a benchmark dataset for multi-label classification.
AudioSet provides balanced and unbalanced training sets.
The balanced set consists of 22 thousand audio clips and the unbalanced set contains around 2 million audio clips.
It also provides 20 thousand balanced audio clips for evaluation (more data statistics can be found in Table~\ref{tab:data_stats} in Appendix~\ref{app:data_stats}). 
\end{itemize}

For each audio clip $\va$, 
we first compute the cosine similarity between it and every possible textual label in the tri-modal representation space. Then we predict the label $t$ with the highest similarity:
\begin{align}
\arg\max_{i} \cos({\vt^{(i)}, \va})\,.
\end{align}

\subsection{Main results}
Our prediction results for AT retrieval are given in Table~\ref{tab:audiocap-retrieval_test} and for zero-shot classification in Table~\ref{tab:zeroshot-clf-merged-prompt} (Appendix~\ref{app:metric} contains qualitative results of the tri-modal representations).

\paragraph{Initializing with visual CLIP weights helps.}
Comparing VA-Rand to \modelname, we see accuracy increases in all classification and retrieval setups. For example, on AudioCaps, \modelname outperforms VA-Rand by 4.5\% R$@$1 and 13.6\% R$@$10. This confirms that the findings of~\citet{AST} carry-over to unsupervised audio pre-training.

\paragraph{Pivoting works well for Audio $\rightarrow$ Text.} \modelname 
exhibits surprisingly strong performance on AT retrieval tasks and zero-shot classification. For example,
it outperforms the supervised baseline~\citep{audio_retrieval} by 2.2\% R$@$1 for text retrieval, without being trained or fine-tuned on Clotho, and without ever having seen an aligned AT pair.

\paragraph{Prompting (usually) helps.} Inspired by the zero-shot image classification setups of CLIP~\citep{CLIP}, we prefix textual labels with a prompt in zero-shot audio classification.
We empirically find that the prompt `\textit{the sound of}' works well.
Using it greatly improves zero-shot multi-class classification accuracy (see Table~\ref{tab:zeroshot-clf-merged-prompt}).
Take \modelname, the prompt gives rise to an improvement of 7.2\% on ESC50 
and 6.9\% on US8K, 
but hurts multi-label classification performance on AS. 

\paragraph{Random curation helps.}
Even when the audio-text pairs used to train that objective are sampled entirely at random (+AT w/ RC),
\modelname improves, e.g., R@1 for Text $\rightarrow$ Audio retrieval increases from 0.8\% to 3.8\%. We conjecture that RC at least makes audio representations aware of and lean towards the text cluster of the joint VT representation space.\footnote{Concretely, VA pre-training pushes audio embeddings towards the image cluster (V) of the VT space of the pre-trained CLIP, but it does not guarantee that audio embeddings will be as close to the text cluster (T) of the VT space as to V. Random curation provides an estimate of the text-cluster's distributional properties, i.e., the audio embeddings are moved on top of the distribution of the text cluster of the VT space \emph{explicitly}; surprisingly, this crude "semantic-free" alignment method improves the quality of audio-text alignment.} While this result also holds for AS classification (+1.5\% mAP), performance decreases for ESC50 (-5.5\% accuracy) and US8K (-2.4\% accuracy).

\paragraph{Unsupervised curation is universally helpful.} \modelname fine-tuned with unsupervised audio captions (+AT w/ AC) outperforms both pivoting (\modelname) and random curation (+AT w/ RC) in all cases. 
Thus, explicitly mining unsupervised AT pairs can be a helpful zero-resource approach. Performance with automatically generated captions (FC) is similar to captions written by humans (AC).

\paragraph{Supervision is still the most helpful.} Fine-tuning \modelname on GC pairs leads to the highest accuracies on ESC50 and US8K.
However, we do not see similar improvements on AS, presumably because multi-label classification is more challenging and requires more direct language supervision, such as audio labels.
This is further evident when we fine-tune \modelname on GL and obtain the highest accuracy (18.9\% mAP) on AS (see Table~\ref{tab:zeroshot-clf-merged-prompt}). 

For retrieval, GL uses only audio labels as the text, which provide less dense language supervision than GC and is thus slightly worse than GC,
but still, it gives better AT alignment than all automatic methods.
As captions become semantically further from the audio-caption domain, e.g., GC < AC < FC < VC, the AT alignment becomes weaker, and thus leading to worse retrieval performance.
The fine-tuned audio encoder generalizes to the out-of-domain Clotho successfully, displaying a trend similar to AudioCaps.

\begin{figure}[t]
	\begin{center}
		\includegraphics[width=0.99\linewidth]{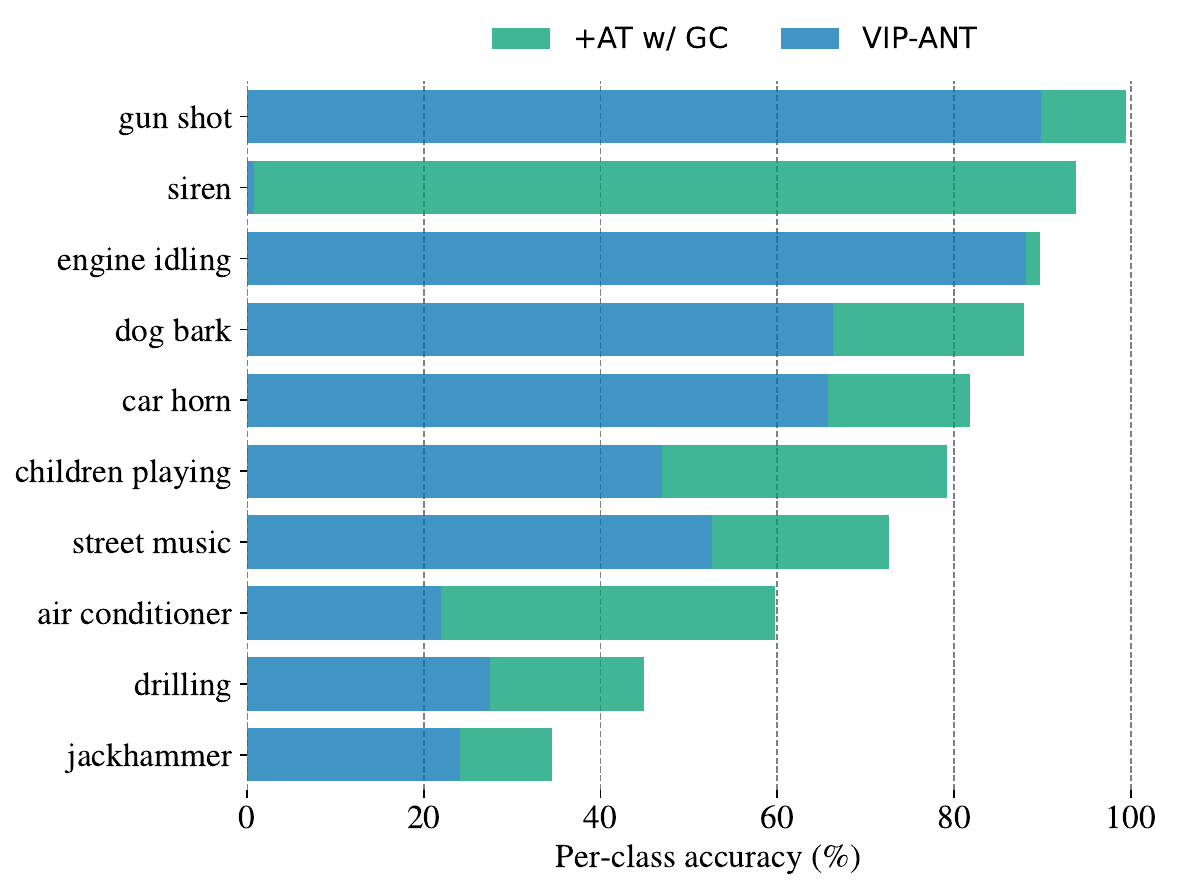}
	\end{center}\vspace{-1em}
	\caption{\label{fig:per-class-acc-us8k}
		Per-class accuracy on US8K.}
	\vspace{-.9em}
\end{figure}

\begin{figure*}[ht]
	\centering
	\begin{subfigure}{.45\linewidth}
		\begin{center}
		\includegraphics[width=.99\linewidth]{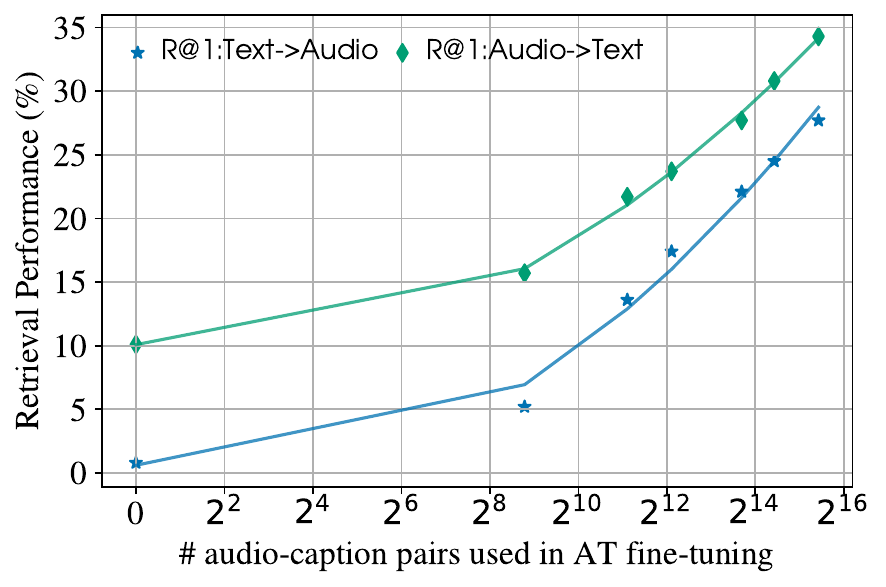}  
		\end{center}
		\vskip -.25em
		\caption{\label{fig:retrieval_level_of_supervision}
	R@1 of AT retrieval on AudioCaps test set.}
	\end{subfigure}
	\hfill
	\begin{subfigure}{.45\linewidth}
	\begin{center}
	\includegraphics[width=.99\linewidth]{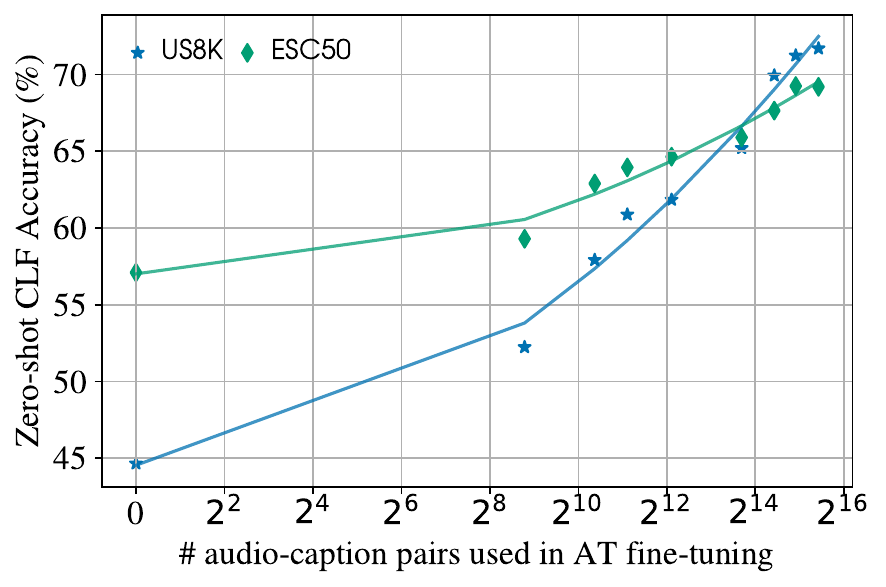}  
	\end{center}
	\vskip -.25em
	\caption{\label{fig:clf_level_of_supervision}
	Zero-shot classification (CLF) on ESC50 and US8K.}
	\end{subfigure}
	\vspace{-.55em}
	\caption{\label{fig:retrieval_clf_level_of_supervision}
	Audio retrieval and zero-shot classification performance versus level of language supervision. 
	}
	\vspace{-.0em}
\end{figure*}

\paragraph{Supervision improves per-class accuracy in general.} We further plot zero-shot classification accuracy for each audio class (see Figure~\ref{fig:per-class-acc-us8k} for US8K and Figure~\ref{fig:per-class-acc-esc50} in Appendix~\ref{app:more_results} for ESC50). Clearly, language supervision improves per-class accuracy in general. The highest improvement is observed on `siren' because `siren' rarely appears in image descriptions while GC contains a lot of textual descriptions of `vehicle' audio.

\subsection{Level of language supervision}
We have observed that AT fine-tuning on AT pairs mined without any additional supervision (e.g., AC, FC, and VC) can improve the AT alignment, but supervised alignments are still the most effective.
But: how much supervised data is really needed?
To understand the relationship between supervision and performance, we vary the number of gold AT pairs (i.e., training samples of AudioCaps) used for AT fine-tuning.
On the audio-text retrieval task (see Figure~\ref{fig:retrieval_level_of_supervision}), unsurprisingly, fine-tuning on more aligned AT pairs results in higher audio-text retrieval / zero-shot classification performance.
Surprisingly, using only 442 (around 1\%) AT pairs of AudioCaps gives rise to as strong AT alignment as VT alignment (\textit{cf.} OracleAV-CLIP in Table~\ref{tab:audiocap-retrieval_test}).

As we increase the number of supervised AT pairs used during fine-tuning, we observe a roughly linear relationship between zero-shot performance and the log of the number of supervised pairs (this observation is similar to \citet{kaplan2020scaling}'s observations regarding Transformers).
While it is not clear how reliable extrapolations from this roughly linear trend are, we roughly estimate the amount of annotated AT pairs required for the zero-shot performance to equal human parity for ESC50 of 81\% \cite{ESC50}: our estimate is that  $2^{21} \approx 2\text{M}$ supervised audio caption pairs would be needed. We are hopeful both (1) that larger curated audio-text datasets will become available; and (2) that future work can improve the data efficiency of the pre-training process.


\section{Conclusion}

We have presented \modelname for unsupervised audio-text alignment induction.
Based on the pivoting idea,
our model learns image-text alignment and image-audio alignment explicitly and separately via bi-modal contrastive pre-training.
The image modality is shared between the two and thus pivots audio and text in the tri-modal embedding space implicitly, without using any paired audio-text data. 
We empirically find that our model achieves strong performance on zero-shot audio-text tasks.
We further strengthen the audio-text alignment by using varying kinds of audio-text supervision.
Experimental results show that even un-aligned audio-caption pairs can help.

\section*{Acknowledgements}
We would like to thank the AI2 Mosaic team for discussions, the AI2 Beaker team for computing support, and the anonymous reviewers for their suggestions. Yanpeng would like to thank Ivan Titov for his comments on the draft. The work was partially supported by the European Research Council (ERC Starting Grant BroadSem 678254), the Dutch National Science Foundation (NWO VIDI 639.022.518), DARPA MCS program through NIWC Pacific (N66001-19-2-4031), DARPA SemaFor program, and Google Cloud Compute.

\bibliography{acl2022}
\bibliographystyle{acl_natbib}

\appendix

\clearpage
\begin{abstract}
This supplementary material includes 
(1) data statistics (\S~\ref{app:data_stats}), (2) hyperparameters of optimizers (\S~\ref{app:hyperparam}),
(3) supervised audio classification (\S~\ref{app:clf_supervised}), (4) interpolating pre-trained position embeddings for Clotho audio-caption retrieval (\S~\ref{app:interp_pos}), (5) comparison between VAT fine-tuning and AT fine-tuning (\S~\ref{app:val_versus_al_finetuning}), (6) a qualitative study of the geometry of the tri-modal embedding space (\S~\ref{app:metric}), 
and (7) additional findings from the audio-text retrieval task (\S~\ref{app:more_results}).
\end{abstract}

\section{Data statistics}\label{app:data_stats}

Table~\ref{tab:data_stats} presents data statistics of all the datasets used in the paper.

\section{Optimizer hyperparameters}\label{app:hyperparam}

Table~\ref{tab:optimizer_setups} presents optimizer hyperparameters used in our learning tasks.

\begin{table}[ht]\small
\centering
{\setlength{\tabcolsep}{0.em} 
\makebox[\linewidth]{\resizebox{\linewidth}{!}{%
\begin{tabular}{
		@{} m{0.01\linewidth} 
		@{} m{0.30\linewidth} 
		@{\hspace{0.25cm}} M{0.125\linewidth} 
		@{\hspace{0.25cm}} M{0.125\linewidth} 
		@{\hspace{0.25cm}} M{0.125\linewidth} 
		@{\hspace{0.25cm}} M{0.125\linewidth} 
	    @{}
	}
	\toprule
	& Hyperparam. & VA & AT & ESC50 & US8K \\
	\midrule
	& Optimizer & \bothmodels{4}{LARS~\citep{LARS}}{olive!5}\\ 
	& Batch size & 432 & 64 & \bothmodels{2}{50}{orange!5} \\
	& Weight decay & \bothmodels{4}{1e-6 }{olive!5} \\
	& LR of weight & \bothmodels{2}{2e-1}{olive!5} & \bothmodels{2}{1e0}{orange!5} \\
	& LR of bias & \bothmodels{2}{4.8e-3}{olive!5} & \bothmodels{2}{2.4e-2}{orange!5} \\
	& Warmup epoch & \bothmodels{4}{10}{olive!5} \\
	& Training epoch & \bothmodels{2}{25}{olive!5} & \bothmodels{2}{50}{orange!5} \\
	\midrule
	\midrule
	& Hyperparam. & \bothmodels{2}{AS balanced}{white}  &  \bothmodels{2}{AS unbalanced}{white}\\
	\midrule
	& Optimizer & \bothmodels{4}{Adam~\citep{Adam}}{olive!5} \\
	& Batch size & \bothmodels{2}{12}{white} & \bothmodels{2}{128}{white} \\
	& Weight decay & \bothmodels{4}{1e-7}{olive!5} \\
	& Learning rate & \bothmodels{4}{5e-5}{olive!5} \\
	& Warmup step & \bothmodels{4}{1000}{olive!5} \\
	& Training epoch & \bothmodels{2}{25}{white} & \bothmodels{2}{5}{white} \\
	& LR scheduler & \bothmodels{4}{MultiStepLR ($\gamma = 0.5$) }{olive!5} \\
	\bottomrule
\end{tabular}}}}
\caption{\label{tab:optimizer_setups}
Hyperparameters of the optimizers used for VA pre-training, AL fine-tuning, ESC50 classification, US8K classification, balanced AS classification, and unbalanced AS classification.
The learning rate (LR) in balanced AS classification is scheduled by epoch: 5, 9, 10, 11, 12 epochs. In unbalanced AS classification it is scheduled by optimization step: 7.5, 15, 20, 25, 35, 40, 45, 50 thousand steps.
}
\end{table}

\section{Supervised audio classification}\label{app:clf_supervised}

To perform supervised audio classification,
we add a classification head (a linear layer) on top of the pre-trained audio encoder.
For \emph{multi-class} classification, the classification head projects the vector representation of an audio clip onto the class space.
We fine-tune the model by minimizing the cross-entropy loss:
\begin{align}
\sum_{i}\log p(y^{(i)} | \va^{(i)})\,,
\end{align}
where $y^{(i)}$ is the gold label of $\va^{(i)}$. For supervised \emph{multi-label} classification, the classification head estimates the likelihood that an audio clip has some textual label.
We thus minimize the per-label binary cross-entropy loss:
\begin{align}
\sum_{i}\sum_{l} \log p(l = 1 | \va^{(i)})\,,
\end{align}
where $l$ enumerates all possible audio labels.

\noindent\textbf{ESC50 and US8K classification.} We initialize the audio encoder from random initialization, CLIP, and \modelname, respectively.
Among them, \modelname performs best. It surpasses random initialization and CLIP on both datasets (see Table~\ref{tab:clf_esc50_and_us8k}).\footnote{
We find that \modelname initialization leads to fast convergence, so it can bring better classification results than other initialization methods with the same number of training epochs.
}
Notably, it outperforms the strong baseline AST-P on ESC50 (+0.1\%), though AST-P has used gold audio labels for supervised pre-training.

\begin{figure}[!t]
	\begin{center}
		\includegraphics[width=0.98\linewidth]{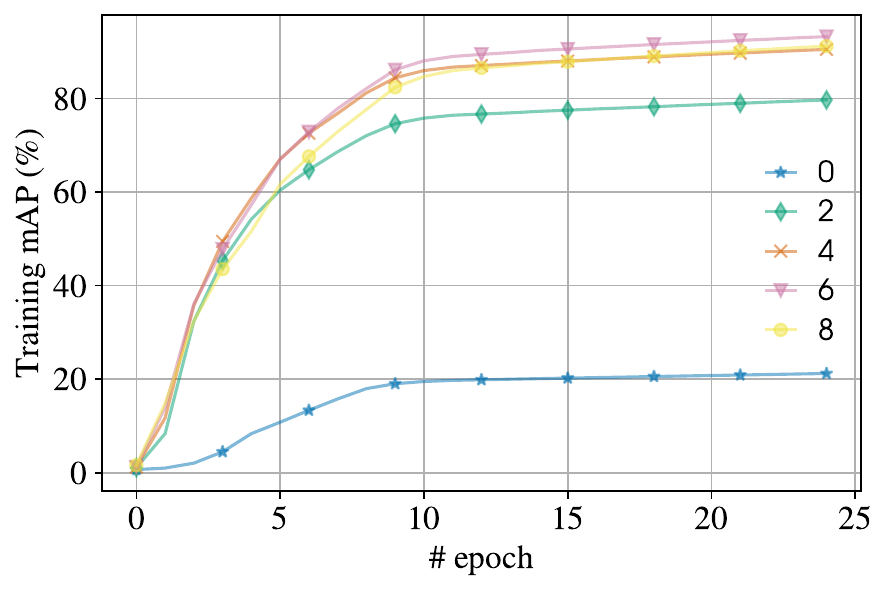}
	\end{center}\vspace{-1em}
	\caption{\label{fig:mAP_tunable_layer}
		Fine-tuning last $k=0, 2, 4, 6, 8$ layers of the pre-trained audio encoder for AS classification. mAP is measured on the AS balanced training set per fine-tuning epoch.}
\end{figure}
\begin{table*}[!t]\small
	\centering
	{\setlength{\tabcolsep}{.5em}
		\makebox[\linewidth]{\resizebox{\linewidth}{!}{%
				\begin{tabular}{llllll}
					{STAT.} & 
					\multicolumn{1}{l}{{AudioSet}} & 
					\multicolumn{1}{l}{{ESC50}} & 
					\multicolumn{1}{l}{{US8K}} &
					\multicolumn{1}{l}{{AudioCaps}} & 
					\multicolumn{1}{l}{{Clotho}}  
					 \\
					\midrule
					\# Train & 2041789 (unbalanced) & 2000 (5-fold) & 8732 (10-fold) & 44118 ($\times 1$ caption) & 3839 (dev-train)  \\
					\# Dev & 22160 (balanced) & & &  & 1045 (dev-val) \\
					\# Val &  &   &   & 441 ($\times 5$ caption) & 1045 (dev-test)\\
					\# Test & 20371 (balanced) &  &   & 860 ($\times 5$ caption) & 1043 (withheld)  \\
					\# Class & 527 & 50 & 10  & & 5 captions / audio \\
					Duration & 10s & 5s  & 0-4s & 10s &  15-30s  \\
					Task & Multi-label CLF & Multi-class CLF & Multi-class CLF & Captioning & Captioning \\
					Source & YouTube & Freesound & Freesound  & YouTube (AudioSet) & Freesound \\
					\bottomrule
	\end{tabular}}}}
	\caption{\label{tab:data_stats}
		Statistics of the data used in this paper. CLF is the abbreivation of "classification". In AudioSet~\citep{audioset} audio clips come from distinct videos. Balanced split means that there are at least 59 samples for each of 527 sound classes. We managed to download 18036 out of 22160 videos in the balanced training split, 16416 out of 20371 videos in the test / evaluation split, and 1715367 out of 2041789 videos in the unbalanced split.
	}
\end{table*}
\begin{table}[ht]
\centering
	{\setlength{\tabcolsep}{0.75em}
	\makebox[\linewidth]{\resizebox{\linewidth}{!}{%
			\begin{tabular}{
					@{\hspace{0.05cm}} p{1.65cm} 
					@{\hspace{0.55cm}} P{1.2cm}
					@{\hspace{0.35cm}} P{1.2cm}
					@{\hspace{0.35cm}} P{1.2cm}
					@{\hspace{0.35cm}} P{1.5cm}
				}
				\toprule
				\multicolumn{5}{c}{AS Classification} \\
				\cdashlinelr{1-5}
				Dataset & AST &  AST$^\star$ & AST$^\dagger$ &  \modelname \\
				\cmidrule{1-5}
				Unbalanced & & & 43.4 & \textbf{44.7} \\
				Balanced  & 34.7 & 35.8 & 31.4 & \textbf{37.9} \\
				\toprule
				\toprule
				\multicolumn{5}{c}{US8K and ESC50 Classification} \\
				\cdashlinelr{1-5}
				Dataset & AST-S &  AST-P & CLIP &  \modelname \\ 
				\cmidrule{1-5}
				US8K  &   &   & 82.5$_{\pm 6.0}$ & \textbf{86.0}$_{\pm 2.8}$\\ 
				ESC50 & 88.7$_{\pm 0.7}$ & 95.6$_{\pm 0.4}$ & 89.7$_{\pm 1.5}$ & \textbf{95.7}$_{\pm 1.4}$ \\
				\bottomrule 
\end{tabular}}}}
\caption{\label{tab:clf_esc50_and_us8k}
Multi-label classification mAPs (\%) on AS and Supervised audio classification accuracies (\%) on ESC50 and US8K. 
AST, AST-S, and AST-P indicates the results reported by~\citet{AST}. We follow their suggestions and test the their best model (AST$^\star$) on our test set. Note that the best model has been trained on the combination of balanced and unbalanced AS training sets. $^\dagger$ indicates that we follow the settings of AST and train it on our data. 
CLIP and \modelname indicate that the audio encoder is initialized from CLIP and from \modelname, respectively.
}
\end{table}

\noindent\textbf{AS classification.}
We consider balanced and unbalanced training for AS classification and train an individual model on the balanced set and the unbalanced set, respectively.
Since the audio encoder has been pre-trained on the unbalanced AudioSet training set, it can be directly used without further fine-tuning.
Nevertheless, we fine-tune the last $k$ layers of the Transformer architecture of \modelname and investigate whether task-specific fine-tuning helps (see Figure~\ref{fig:mAP_tunable_layer}).
When $k=0$ the model is basically a linear probe. It inspects if contrastive pre-training learns separable audio representations.
As we increase $k$, i.e., fine-tuning more layers, the model exhibits a tendency of over-fitting the training set.
We use $k=4$ as a trade-off between under-fitting and over-fitting. Our model achieves the best mAP of 37.9\% for balanced training, which surpasses AST by 6.5\% (see Table~\ref{tab:clf_esc50_and_us8k}).
While for unbalanced training, we find it crucial to fine-tune the whole model.
Again, our model outperforms AST (+1.4\% mAP).

\section{Position embedding interpolation}\label{app:interp_pos}

Clotho~\citep{audiocaps-clotho} audio has a duration of 15-30 seconds, which is longer than 10-second audio clips used in pre-training. To apply our pre-trained audio encoder to Clotho audio-caption retrieval, we up-sample the pre-trained positional embeddings to account for the longer audio token sequences. Table~\ref{tab:clotho_interp_pos_test} shows retrieval performance of 10-second audio and 18-second audio. In general, longer audio gives rise to better audio-caption retrieval performance.

\begin{table*}[t]\small
	\centering
	{\setlength{\tabcolsep}{1.2em} 
		\makebox[\linewidth]{\resizebox{\linewidth}{!}{%
				\begin{tabular}{p{0.01cm} p{2cm} rrrrrrrr}
					\toprule
					& \multirow{3}{*}{Model} & \multicolumn{4}{c}{10-second Clotho (eval)} & \multicolumn{4}{c}{18-second Clotho (eval)} \\
					\cmidrule{3-10}
					& & \multicolumn{2}{c}{Text$\rightarrow$Audio} & \multicolumn{2}{c}{Audio$\rightarrow$Text} &  \multicolumn{2}{c}{Text$\rightarrow$Audio} & \multicolumn{2}{c}{Audio$\rightarrow$Text}\\ 
					& & {\scriptsize R$@$1} & {\scriptsize R$@$10} & 
					{\scriptsize R$@$1} & {\scriptsize R$@$10} &   
					{\scriptsize R$@$1} & {\scriptsize R$@$10} & 
					{\scriptsize R$@$1} & {\scriptsize R$@$10} \\ 
					\midrule
					\multirow{6}{*}{\rotatebox[origin=c]{90}{{\footnotesize Zero-resource}}}
					& VA-Rand & 1.4 & 7.4 & 3.2 & 13.1 & 1.3 & 7.5 & 3.2 & 13.5 \\
					& \modelname & 1.9 & 10.1 & 6.1 & 23.7 & 1.9 & 9.5 & 7.0 & 25.6\\
					&  +AT w/ AC     & 5.9 & 26.3 & 8.2 & 30.3  & 6.7 & 29.1 & 7.1 & 30.7\\
					&  +AT w/ FC     & 5.7 & 26.6 & 6.6 & 28.0 & 6.5 & 27.7 & 7.8 & 29.7    \\
					&  +AT w/ VC     & 5.2 & 25.2 & 7.0 & 25.9  & 5.5 & 25.6 & 7.6 & 28.2 \\					
					&  +AT w/ RC     & 3.5 & 16.3 & 5.7 & 23.6 & 3.5 & 16.9 & 5.5 & 24.9 \\
					\cdashlinelr{2                                                                                                                                                                                                                                                                                                                                                                                                                                                                                                                                                                                                                                                                                                                                                                                                                                                                                                                                                                                                                                                                                                                                                                                                                                                                                                                                                                                                                                                                                                                                                                                                                                                                                                                                                                                                                                                                                                                                                                                                                                                                                                                                                                                                                                                                                                                                                                                                                                                                                                                                                                                                                                                                                                                                                                                                                                                                                                                                                                                                                                                                                                                                                                                                                                                                                                                                                                                                                                                                                                                                                                                                                                                                                                                                                                                                                                                                                                                                                                                                                                                                                                                                                                                                                                                                                                                                                                                                                                                                                                                                                                                                                                                                                                                                                                                                                                                                                                                                                                                                                                                                                                                                                                                                                                                                                                                                                                                                                                                                                                                                                                                                                                                                                                                                                                                                                                                                                                                                                                                                                                                                                                                                                                                                                                                                                                                                                                                                                                                                                                                                                                                                                                                                                                                                                                                                                                                                                                                                                                                                                                                                                                                                                                                                                                                                          -10}
					\multirow{2}{*}{\rotatebox[origin=r]{90}{\parbox{.3cm}{Zero-shot}}}
					&  +AT w/ GL     & 6.0 & 27.1 & 6.1 & 25.4 & 6.7 & 29.0 & 6.8 & 27.0\\	
					&  +AT w/ GC     & 10.2 & 39.0 & 10.3 & 37.2 & \textbf{11.1} & \textbf{40.5} & \textbf{11.8} & \textbf{41.0}  \\	
					\bottomrule
	\end{tabular}}}}
	\caption{\label{tab:clotho_interp_pos_test}
		Interpolating positional embeddings to account for Clotho audios which are longer than 10 seconds.
	}
\end{table*}

\section{VAT versus AT fine-tuning}\label{app:val_versus_al_finetuning}

\begin{figure*}[t]
	\centering
	\begin{subfigure}{.48\linewidth}
		\centering
		\includegraphics[width=.99\linewidth]{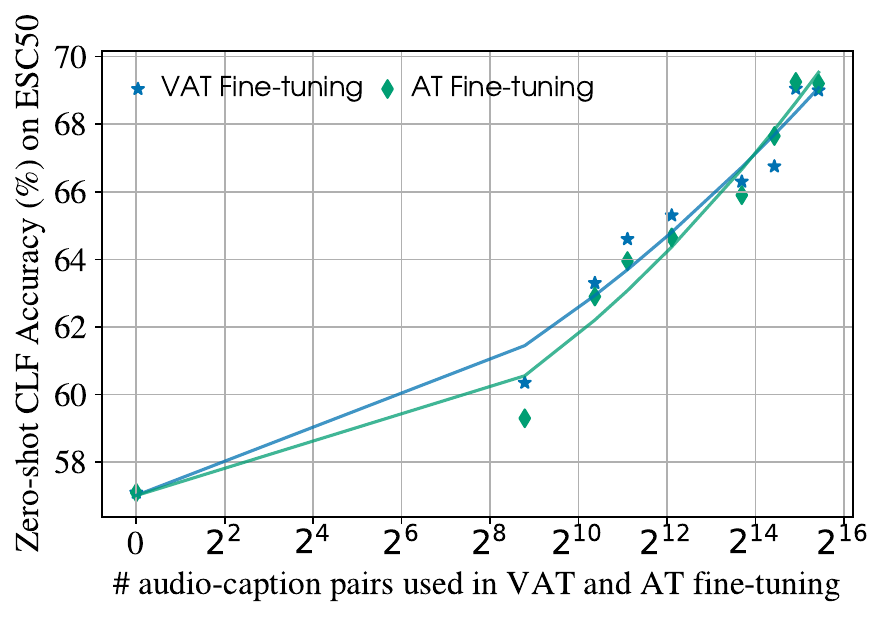}  
		\vskip -.25em
		\caption{\label{fig:val_vs_al_clf_level_of_supervision}
	Zero-shot classification (CLF) accuracy versus level of supervision.}
	\end{subfigure}
	\hfill
	\begin{subfigure}{.48\linewidth}
	\centering
	\includegraphics[width=.99\linewidth]{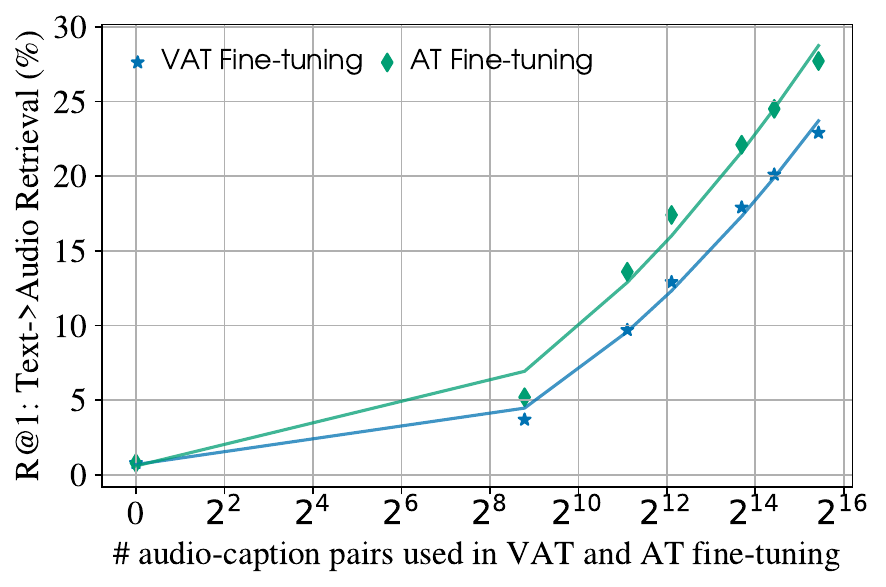}  
	\vskip -.25em
	\caption{\label{fig:val_vs_al_retrieval_level_of_supervision}
	R@1 of audio retrieval with text as the query versus level of supervision.}
	\end{subfigure}
	\vspace{-.55em}
	\caption{\label{fig:val_vs_al}
	Comparing VAT and AT fine-tuning on zero-shot ESC50 classification and AudioCaps audio retrieval.}
\end{figure*}

Given caption-augmented AudioCaps audio~\citep{audiocaps}, we can improve the pre-trained audio encoder via contrastive vision-audio-text (VAT) fine-tuning and contrastive audio-text (AT) fine-tuning. Figure~\ref{fig:val_vs_al} shows a comparison between the two fine-tuning techniques on zero-shot ESC50 classification and AudioCaps audio retrieval.
In general, AT fine-tuning results in better results on the two tasks.

\section{Analyzing tri-modal representations}\label{app:metric}

To better understand the geometry of tri-modal embeddings of our pivoting, unsupervised curation, and supervised curation, we study how AT fine-tuning influences the tri-modal representation space. Specifically, we analyze \modelname (pivoting), \modelname+AT (w/ RC) (unsupervised curation), and \modelname+AT (w/ GC) (supervised curation) using \emph{pivotability}.

\begin{center}
\resizebox{0.8\linewidth}{!}{%
\begin{tikzpicture}[node distance=0mm]

\node[circle, fill=mygreen!35,draw=lightgray!75,anchor=north west,minimum width=1.5cm,minimum height=1.5cm] (A) at (0.5, 4) {\textcolor{black!75}{\bf Audio}};

\node[circle, fill=myorange!35,draw=lightgray!75,anchor=north west,minimum width=1.5cm,minimum height=1.5cm] (V) at (4, 5) {\textcolor{black!75}{\bf Image}};

\node[circle, fill=mypurple!35,draw=lightgray!75,anchor=north west,minimum width=1.5cm,minimum height=1.5cm] (L) at (7.5, 4) {\textcolor{black!75}{\bf Text}};

\node[rounded corners, align=center, anchor=north west, font=\Large, 
minimum width=32mm] (ENCODER) at ([shift=({-1.6cm,-0.1cm})]V.south){\textcolor{gray}{Retrieval Path}};



\draw [-latex,line width=.35mm,black!65] (A) to [bend left] node[above,font=\fontsize{14pt}{16pt}\selectfont]{top-$k$} (V);
\draw [-latex,line width=.35mm,black!65] (V) to [bend left] node[above,font=\fontsize{14pt}{16pt}\selectfont]{top-5} (L);
\draw [-latex,line width=.35mm,black!65] ([xshift=-1.5cm]A.west) -- node[above,font=\fontsize{14pt}{16pt}\selectfont]{start}  (A.west);
\end{tikzpicture}}
\end{center}

Pivotability measures how likely images can pivot audio and text. 
We quantify it for each aligned VAT triplet via a two-step retrieval probe. Starting at a given audio clip, we retrieve $k$ nearest image neighbors; for each image neighbor, we retrieve the top-5 nearest captions. Since each audio clip has 5 gold captions, we compute pivotability as the ratio of the number of retrieved gold captions to 5.
A gold caption may be retrieved more than one time, but we always count it as 1, so pivotability is always between 0 and 1.

\begin{figure}[!t]
	\centering
	\begin{center}
	\includegraphics[width=.99\linewidth]{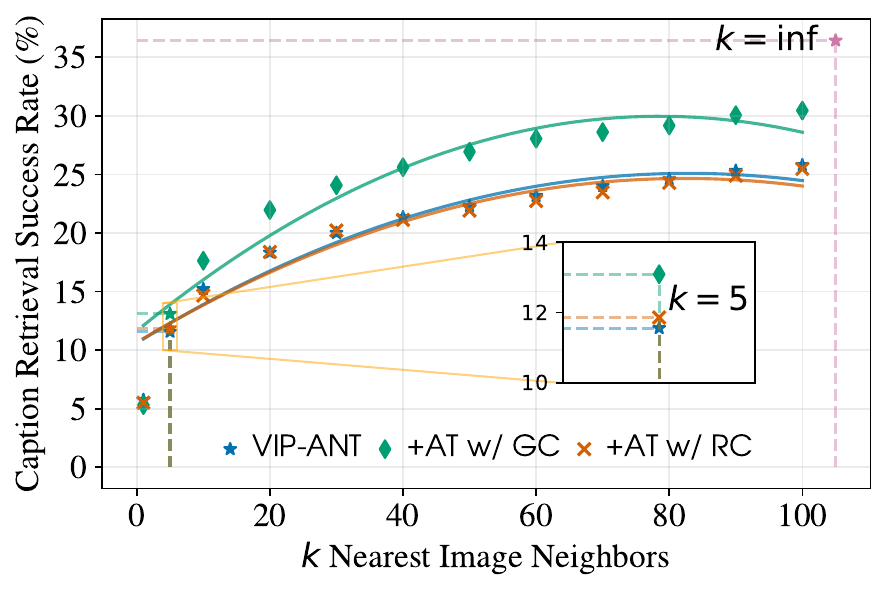}  
	\end{center}
	\vskip -.25em
	\caption{\label{fig:image_pivot} Tri-modal pivotability. +AT (w/ GC) and +AT (w/ RC) indicate that \modelname is further fine-tuned on GC and RC, respectively. 
	}
\end{figure}

We conduct this experiment on AudioCaps test set.
For each $k$, i.e., how many images will be retrieved for a given audio clip, we average pivotability scores over all test triplets (see Figure~\ref{fig:image_pivot}).

\begin{figure}
	\begin{center}
		\includegraphics[width=0.9\linewidth]{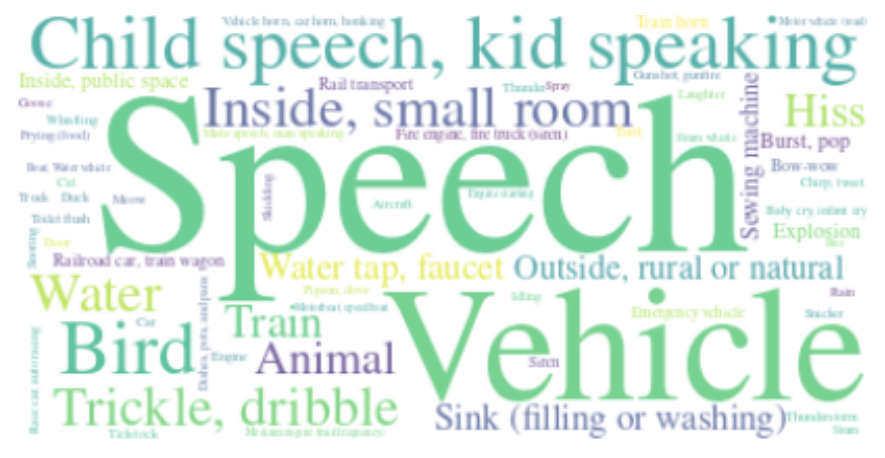}
	\end{center}\vspace{-1em}
	\caption{\label{fig:pivotable_audio} Categories of the audio that can be pivoted with text by images.
	Larger text indicates that the related audio is more likely to be pivoted with text.
	}
\end{figure}

\paragraph{Which pairs are pivotable?}
To study what kinds of audio are more likely to be pivoted with text by images,
we set $k=5$, i.e., 5 images will be retrieved for each given audio clip.
We consider an AT pair as pivotable if at least 3 out of 5 gold captions of the audio clip are retrieved, i.e., pivotability is equal to or larger than 0.6.
Figure~\ref{fig:pivotable_audio} illustrates the categories of the audio clips in pivotable AT pairs.
Unsurprisingly, audio about speech and vehicle is more pivotable because the two categories are among the top three frequent categories in AS.\footnote{Music is the second most frequent category in AS. It is not shown in the figure because AudioCaps excludes all music audio.}
Given that AT fine-tuning improves Audio $\rightarrow$ Image retrieval, we wonder if it could also help find novel categories of audio that can be pivoted with text. 
We find that this is indeed the case (see Table~\ref{tab:novel_categories_pivotable_audio}).
For example, \modelname+AT (w/ GC) finds more fine-grained speech categories because most AT pairs in AudioCaps are about speech.
In contrast, \modelname+AT (w/ RC) finds two additional novel insect categories, presumably because RC suffers from less data bias than GC.

\begin{table}[ht]\small
	\centering
	{\setlength{\tabcolsep}{0.em} 
	\makebox[\linewidth]{\resizebox{\linewidth}{!}{%
	\begin{tabular}{@{} m{0.2\linewidth} @{\hspace{0.35cm}} m{0.6\linewidth} @{} m{0.01\linewidth}}
	\midrule
	+AT w/ GC & `female speech, woman speaking', `narration, monologue', `vibration' &\\
	\midrule
	+AT w/ RC & {\color{orange}{`bee, wasp, etc.'}}, `female speech, woman speaking', {\color{orange}{`insect'}}, `narration, monologue', `vibration' &\\
	\bottomrule
	\end{tabular}}}}
	\caption{\label{tab:novel_categories_pivotable_audio}
		Comparing against \modelname, the two fine-tuned versions of \modelname find novel audio categories in pivotable AT pairs.
	}
\end{table}




\section{Additional results}\label{app:more_results}
\paragraph{Asymmetric audio-text retrieval performance.} 
For Text $\rightarrow$ Audio retrieval, our unsupervised pivoting model is not as good as on Audio $\rightarrow$ Text.
This could be because audio is intrinsically more difficult to retrieve with specificity than text in our corpus, e.g., because sound events co-occur (a baby may cry in street with sirens in the background or in a room with dogs barking), there may be a broader range of captions that accurately describe them.
However, it could also be the case that AT alignment is bounded by VT alignment because VA pre-training biases audio representations towards image representations.
We check this hypothesis by conducting image-text retrieval on AudioCaps. AudioCaps provides aligned image-audio-text triplets, so we simply replace audio with the corresponding image. 
We find that the Text $\rightarrow$ Image retrieval performance of CLIP is much better than the Text $\rightarrow$ Audio retrieval performance of \modelname (see the OracleAV-CLIP row of Table~\ref{tab:audiocap-retrieval_test}).
It is also close to the 
Image $\rightarrow$ Text retrieval performance of CLIP.
In contrast, \modelname exhibits a large gap between the Text $\rightarrow$ Audio retrieval performance and the Audio $\rightarrow$ Text retrieval performance.

\begin{figure}[!t]
	\begin{center}
		\includegraphics[width=0.99\linewidth]{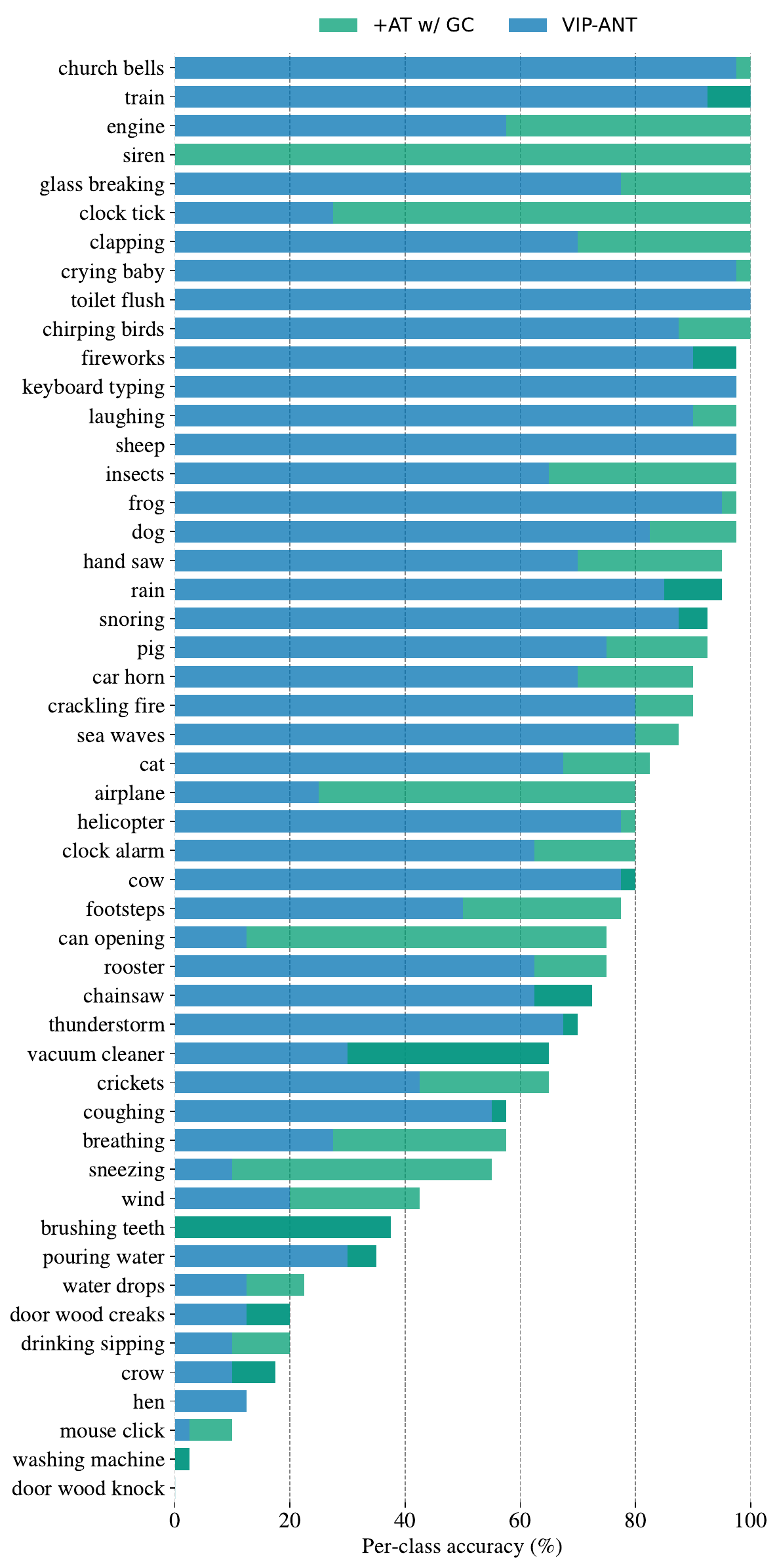}
	\end{center}\vspace{-1em}
	\caption{\label{fig:per-class-acc-esc50}
		Per-class accuracy on ESC50.}
\end{figure}

\paragraph{Per-class accuracy on ESC50} is illustrated in Figure~\ref{fig:per-class-acc-esc50}.

\end{document}